\documentclass[fleqn,10pt]{wlscirep}
\usepackage[utf8]{inputenc}
\usepackage[T1]{fontenc}
\usepackage{bm}
\usepackage{graphicx}
\usepackage{amsmath}
\usepackage{physics}
\usepackage{amsfonts}
\usepackage{amssymb}
\usepackage{ulem}
\usepackage{latexsym}
\usepackage{wasysym}
\usepackage{tcolorbox}
\usepackage{tikz}
\usepackage{varwidth}
\usepackage{capt-of}

\usepackage{mathptmx}

\title{Supersolidity in 
ultra-cold dipolar  gases}

\author[1,*]{Alessio Recati}
\author[1,+]{Sandro Stringari}
\affil[1]{Pitaevskii BEC Center, CNR-INO and Dipartimento di Fisica, Università di Trento, Via Sommarive 14, I-38123 Trento, Italy
}

\affil[*]{e-mail: alessio.recati@cnr.it}
\affil[+]{e-mail: sandro.stringari@unitn.it}

\begin{abstract}
Can a gas behave like a crystal? Supersolidity is an intriguing and challenging state of matter which combines key features of superfluids and crystals. Predicted a long time ago, its experimental realization has been recently achieved in Bose-Einstein condensed (BEC) atomic gases inside optical resonators,  spin-orbit coupled BEC's and atomic gases interacting with long range dipolar forces. The activity on dipolar gases  has been particularly vibrant in the last few years. This perspective article summarizes the main experimental and theoretical achievements concerning supersolidity in the field of dipolar gases, like the  observation of the density modulations  caused by the spontaneous breaking of translational invariance, the effects of coherence and the occurrence of novel Goldstone modes. A series of important issues for the future experimental and theoretical research are outlined  including, among others,  the possible realization of quantized vortices inside these novel crystal structure, the role of dimensionality, the characterisation of the crystal properties and the nature of the phase transitions. 

At the end a brief overview on some other (mainly cold atomic) platforms, where supersolidity has been observed or where supersolidty is expected to emerge is provided.

\end{abstract}
\begin{document}

\flushbottom
\maketitle
\newpage
\tcbset{colback=cyan!5!white,colframe=cyan!75!black,fonttitle=\bfseries}

\begin{tcolorbox}[every float=\centering, title={Superfluid, Supersolid and Crystal phases in ultra-cold dipolar gases}]
    \begin{minipage}[t]{\linewidth}
    \vspace*{0pt}
  \includegraphics[width=\linewidth]{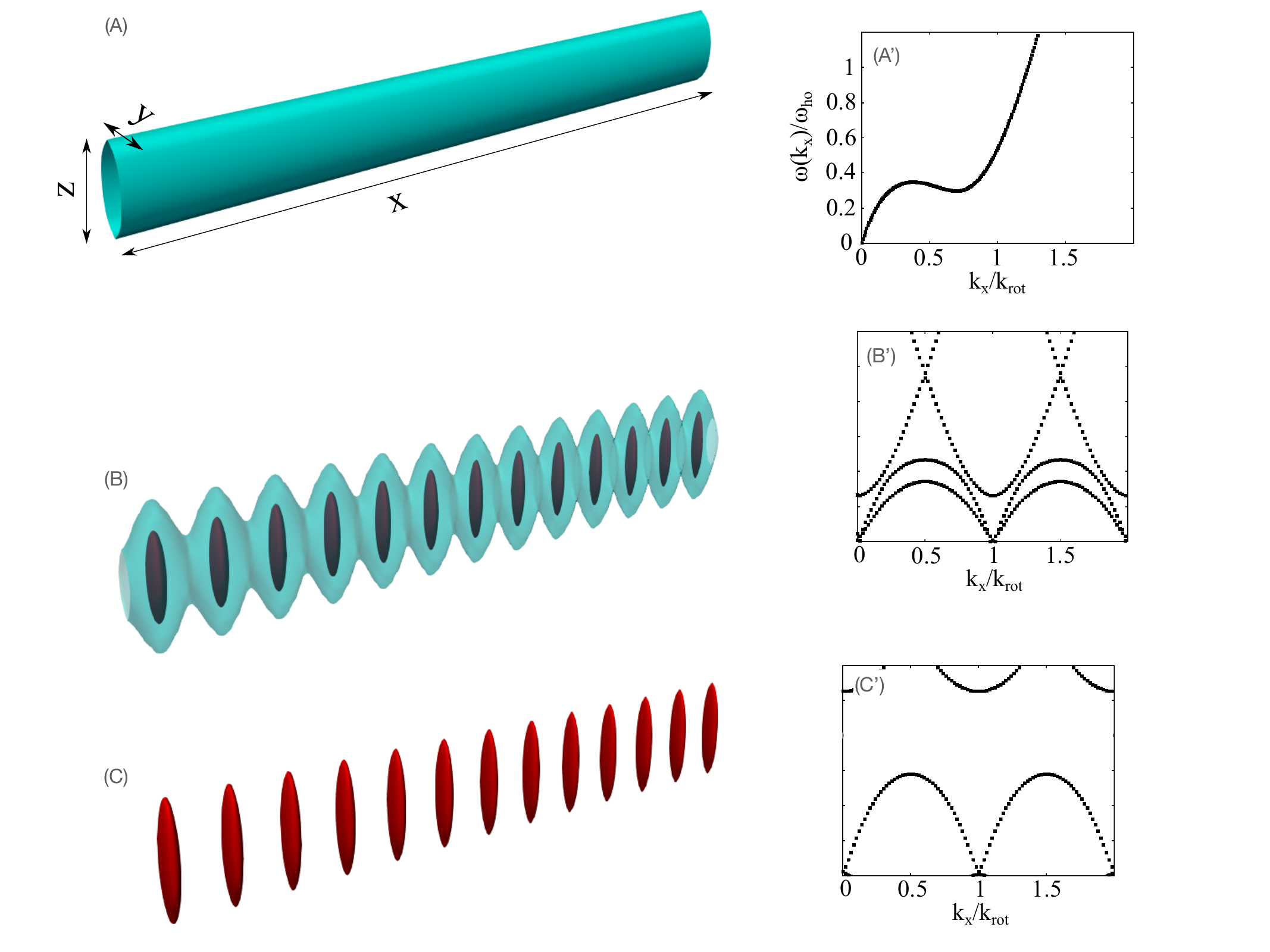}
  
A confined dipolar gas can undergo a transition from a standard homogeneous Bose-Einstein condensed gas (A) to a state where the density is modulated. For not too strong dipolar interactions the system remains fully coherent yielding supersolidity (B) . Increasing the dipolar interaction leads to a system formed by independent self-bound droplets forming a crystal (C). 

The dispersion relation of the system reflects its superfluid and crystal nature : (A') In the homogeneous superfluid the low-momentum dispersion relation is gapless and phononic due to the zero-energy cost of changing the phase of the condensate wave-function and to the stiffness introduced by the interaction. The spectrum shows a roton-like minimum due to the long-range and anysotropic nature of the dipolar potential. 
(B') In the supersolid modulated state, in addition to  the superfluid mode, a novel crystal phonon spectrum appears.  (C') In the indipendent droplet state only the crystal mode is present. In both (B) and (C) phases the excitation spectrum exhibits a typical band structure.

For the sake of clarity an infinitely long one-dimensional structure is shown here. 

    \end{minipage}\hfill%
\end{tcolorbox}

\section*{Introduction}

Despite first being predicted back in the 1960s (refs.\cite{GrossAP1960,AndreevLifshitz}) and 1970s (refs~\cite{chester1970,Leggett1970}), supersolidity is now one of the most active research frontiers in the field of ultracold atoms. In this intriguing quantum phase of matter  both superfluid and crystalline
properties coexist as a consequence of the simultaneous breaking of phase and translational symmetry. 
Supersolidity is in general the result of interactions, which, under peculiar and special conditions, are able to ensure the co-existence of coherence, superfluidity and crystallization, giving rise to a new spectrum of elementary excitations (see Box).


After the unsuccessful experimental attempts in solid helium \cite{Chan2}(see also Ref.~\cite{Balibar}) it emerged clearly that supersolidity is more likely expected to occur when the modulations of the density, corresponding to the spontaneous breaking of translational symmetry,  involve a large number of particles and not just single atoms as it would happen in traditional solids \cite{nepo1971,Hydro-Rica2007,Boninsegni2012}.  Relevant examples are coherent droplets or stripes, where each droplet or stripe contains a mesoscopic number of atoms. For this reason such states can be named cluster supersolids.
Supersolidity was actually first experimentally realized in Bose–Einstein condensates (BECs) with spin–orbit coupling \cite{Li2017} where supersolidity takes the form of stripes, inside optical resonators \cite{Leonard2017} and in  Bose gases interacting with long-range dipolar forces~\cite{Pisa1,Stut1,IBK1}, where it takes the form of periodically arranged large clusters. This perspective article focuses on \sout{long-range} dipolar  gases, where compelling  theoretical predictions and important experimental achievements have emerged in the recent years.

\section*{Background}
A typical scenario for the occurrence of supersolidity is the existence of a roton-like spectrum -- a spectrum with a local minimum (roton gap) at a finite momentum $k_{\textrm{rot}}$ -- in the superfluid (homogeneous) phase. A small enough roton gap can trigger an instability towards a state characterised by a structure with period $\propto k_{\textrm{rot}}^{-1}$ (see also Box 1). The observation that the presence of negative Fourier components in the two-body interaction  can lead to a a rotonized spectrum and eventually to a periodically-modulated condensate state in interacting bosons \cite{GrossAP1960}, has been fundamental in the search for a supersolid phase in ultracold gases. Other model calculations  of bosons interacting with soft core potentials have focused on this new phase of matter either using mean field~\cite{Pomeau93,Pomeau94} and ab-initio Monte Carlo calculations \cite{Boninsegni_SS2012}.

 \subsection*{Precursor phenomena of supersolidity}

Even in the superfluid (non supersolid) phase, dipolar Bose gases exhibit important differences with respect to usual Bose-Einstein condensed gases interacting solely with short range interactions (see the recent review \cite{Chomaz_Review2023}). This is due to the long range and anisotropic nature of the dipole force and its consequent tendency to favor instability and collapse if atoms are allowed to align head-to-tail along the direction of the external magnetic field. This tendency to collapse can be counteracted by imposing strong enough trapping confinement along the orientation of the dipole moment. In such trapping conditions the long range nature of the dipolar interaction is responsible for the occurrence of a rotonic structure in the excitation spectrum\cite{SantosRoton,Odell2003}. The occurrence of the roton minimum and the corresponding sharp peak exhibited by the static structure factor  and the static response \cite{Bisset2019} reflects the intrinsic tendency to solidification exhibited by the system, in analogy with the  transition to the solid phase exhibited by liquid $^4$He \cite{celli1972,balibar2006}.

\begin{figure}[h]
\centering
\includegraphics[width=0.5\linewidth]{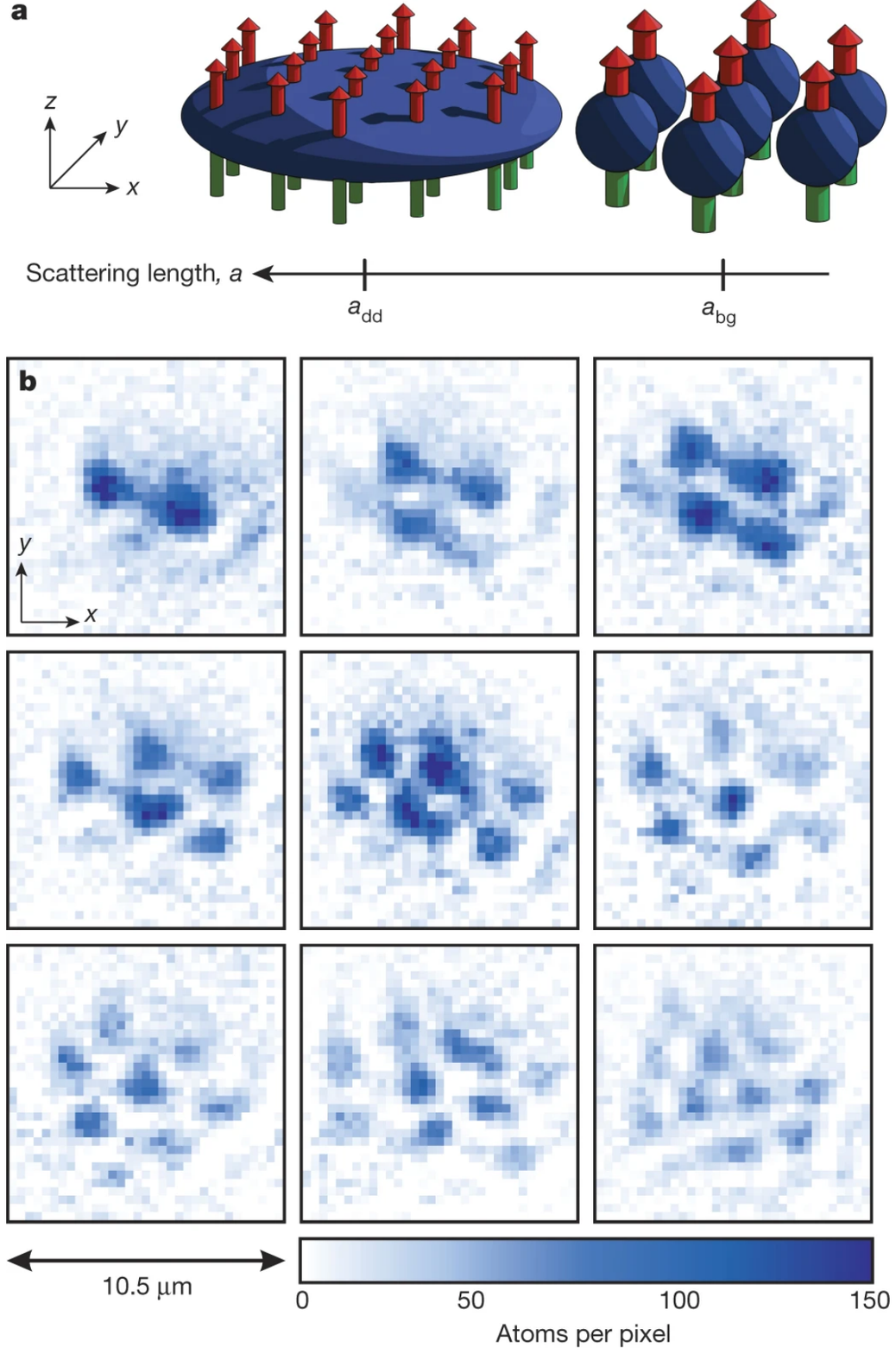}
\caption{Observation of crystallization of droplets in a superfluid dipolar gas. \textbf{a}: Schematic of the experimental procedure: A stable, strongly dipolar BEC of dysprosium atoms in a pancake-shaped trap is prepared (left). By decreasing the scattering length $a$, an instability (a.k.a. Rosensweig instability) is  induced. Following this instability, the atoms clustered to droplets in a triangular pattern (right). \textbf{b}: Representative single samples of droplet patterns imaged in situ, with droplet numbers ranging from two to ten. [Adapted from Ref.~\cite{holger2016}] }
\label{fig:dropspfau}
\end{figure}

A first important step was the understanding of the crucial role played by the beyond mean field corrections -- also known as Lee-Huang-Yang, shortly LHY, corrections\cite{LHY1957} -- to the equation of state \cite{FischerLHY,Pelster},  which, in spite of their smallness, can play a crucial role in stabilizing the system \cite{PetrovDrop,Wachter2016}, avoiding the collapse predicted by mean field theories, when the role of the attractive component of the force encompasses a critical value. A spectacular consequence, caused by beyond mean field effects, was the observation of self bound droplets (see Fig.~\ref{fig:dropspfau}), both in dipolar gases \cite{BarbutDrop2016,holger2016,Schmitt2016,ChomazDroplet2016} as well as in binary mixtures of BEC gases with attractive interparticle interaction \cite{FattoriDrop,TarruellDrop,TarruellDrop2,FattoriColl}.  The latter systems do not however exhibt a rotonic structure and consequently are not expected to show  supersolid features.

\begin{figure}[h]
\centering
\includegraphics[width=0.75\linewidth]{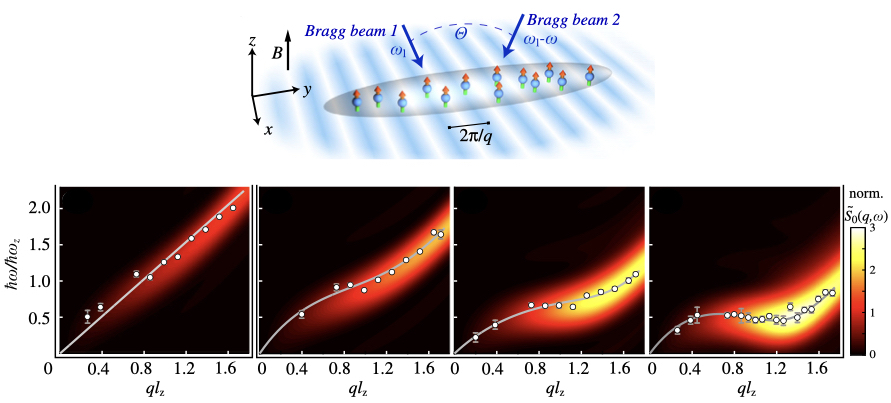}
\caption{Probing the existence of the roton mode and its softening using Bragg spectroscopy for a gas cloud of tens of thousands $^{166}$Er atoms schematically represented in the upper panel. In the lower panel the experimentally extracted spectra (white dots) are compared with the results obtained within Bogoliubov theory (solid line). The color scale represent the dynamic structure factor, which, as expected, shows a strong enhancement when the roton softens. [Adapted from Ref.~\cite{petter-roton2019}]}
\label{fig:rotonibk}
\end{figure}

A further major step towards the realization of supersolidity was the experimental observation in an elongated cloud~\cite{IBKrotons,petter-roton2019} of the predicted rotonic excitation of a dipolar BEC gas and the roton softening followed by the roton collapse, when the value of the s-wave scattering length is tuned to small enough values (see Fig. \ref{fig:rotonibk}).
Signatures of the radial and angular roton excitations in two dimensional configurations and of their softening around a droplet crystallization transition have been also recently found \cite{Roton2DPfau2021}.
Finally, in the meantime, theorists were able to identify a tiny region of the relevant parameters in the phase-space diagram, where beyond mean field effects, are responsible for stabilizing coherent arrays of quantum droplets, yielding the long sought conditions for the realization of supersolidity in dipolar BEC gases \cite{Roccuzzo1}.

\subsection*{Experimental evidence of supersolidity}


In the year 2019 the field was ready for a major breakthrough: three experimental groups reported on the evidence of coherence between dipolar droplets~\cite{Pisa1,IBK1,Stut1}. As theoretically predicted~\cite{Roccuzzo1}, the range of $\epsilon_{dd}$ in which coherence persists is very small.
However it was clear that the system shows three distinct phases: a superfluid state, a coherent droplet ({\it supersolid}) state and an incoherent droplet ({\it crystal}) state (see Box).  The number of droplets in these first experiments was limited to 2-4
in a linear configuration and the names supersolid or crystal are suggestive of the hypothetical bulk system corresponding to the thermodynamic limit. Presently larger chains, as well as two dimensional supersolid-like configurations have been realised in ultra-cold dipolar gases \cite{Norcia2021_2D,Bland2022_2D}. 

After the realisation of a spontaneously modulated BEC and the experimental proof of its coherence the efforts of experimentalists and theorists were devoted to get additional signatures of the supersolid nature of these systems  and in particular  to explore the different kinds of sound propagating in a dipolar supersolid. The presence of both crystalline order (standard long-range order) and superfluid order (off-diagonal long-range order), corresponding to the  spontaneous breaking of  continuous translational invariance and of the U(1) symmetry (particle conservation), respectively,  is in fact   expected  to give rise to a novel series of Goldstone modes. 
Simulations on bulk atomic Bose gases interacting with soft-core potentials \cite{Pomeau93} -- employing Monte-Carlo techniques \cite{Boninsegni_SS2012} and  mean-field calculations \cite{Macri_SS,Ancilotto2013}  have confirmed the predictions of the Goldstone theorem and characterised the proper dispersion relation of the excitations  in the three distinct phases,  i.e., a single gapless mode in the superfluid phase, $d$ modes in the crystal phase and $d+1$ modes in the supersolid phase, where $d$ is the number of dimensions in which  traslational invariance is broken. It is important to notice that the presence  of $3+1$ phonon modes was already predicted in the seminal paper by Andreev and Lifshitz \cite{AndreevLifshitz} for the hypothetical supersolid phase of $^4$He. In the case of  a highly elongated linear supersolid the dispersion relation is expected to include a single gapless mode in the superfluid  and in the crystal incoherent droplet phases, and two gapless modes in the supersolid phase (see Box ``Superfluid, Supersolid and Crystal phases in ultra-cold dipolar gases'').  As already pointed out by Andreev and Lifshitz the sound modes are hybridized, i.e,  are neither purely crystal nor purely superfluid modes. In the case of a small superfluid density, the modes are expected to be essentially decoupled \cite{AndreevLifshitz}.

\begin{figure}[h]
\centering
\includegraphics[width=0.5\linewidth]{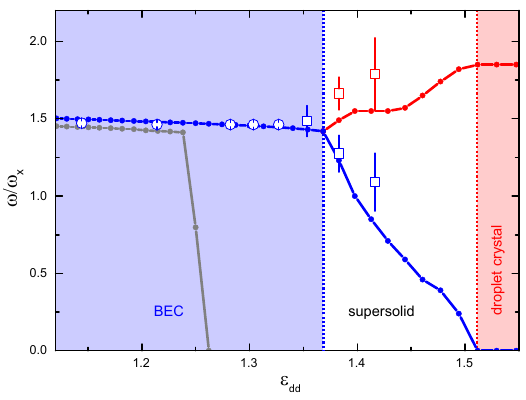}
\caption{Longitudinal breathing mode frequency $\omega$ of a dipolar gas in the superfluid (BEC, blue-shaded region), supersolid (SS, white region) and crystal (red-shaded region) phases. The frequency is renormalised with respect to the longitudinal trapping frequency $\omega_x$. Filled points and continuos lines are the numerically extracted values, circles and squares the experimental values. The grey line show the theoretical result for a BEC in absence of the Lee-Huang-Yang correction. [Adapted from Ref.\cite{Pisa2}]}
\label{fig:breathing}
\end{figure}

Quite remarkably,  just a few months after the first evidence of the supersolid state, the Goldstone modes of supersolid dipolar gases were addressed in a series of joint experimental and theoretical works \cite{Pisa2,Stut2,NatNews2019,Ibk2}. 
In harmonically trapped finite size systems -- which is the  typical situation of the presently available configurations -- the continuous dispersion relation takes the form of a series of discretised energy levels.   As a concrete example let us consider the response of a highly elongated system system to a longitudinal compression. The gas starts breathing and while in the superfluid and in the incoherent droplet phases the motion of the cloud is characterised by a single frequency, in the supersolid phase two frequencies can be instead clearly identified, as can be seen in Figure \ref{fig:breathing}. The eigenmodes of such frequencies can be shown to be dominated either by the relative motion of the droplets or by the change in amplitude of the droplets without a change in the position of the peaks. For completeness in Fig. \ref{fig:breathing} the theoretical results for the mean-field equation of state is also reported as a grey line. The crucial role played by beyond mean field contribution to the equation of state is clear. In the absence of such a term 
one  would predict a collapse before reaching the transition from the superfluid to the density modulated phase.

As mentioned above an important recent breakthrough in the field of dipolar gases has been the realization of two-dimensional supersolid structures, such as  zig-zag and hexagonal configurations \cite{2DSS-NatureIBK,biagioni2022,Bland2022_2D}, paving the way for future investigations concerning the rotational properties and the role of dimensionality.  The rotational properties of these 2D-like configurations have been already the object of first experimental and theoretical investigations.  
In particular  it has been shown that  the link between the frequency of the  scissors mode, which is naturally excited by suddenly rotating the confining harmonic trap, and the value of the moment of inertia is not trivial \cite{Norcia2021_scissors}, being  strongly affected by the occurrence of additional  low frequency rotational oscillations of the droplets\cite{Roccuzzo2021_inertia}.  This  differs from the case of elongated supersolid configurations, where the droplets are aligned in a 1D like geometry and the excitation of the scissors mode provides direct access to the value of the moment of inertia and to its deviations from the classical rigid value \cite{roccuzzo2, ScissorPisaSS}, in analogy with  usual BEC gases \cite{Odelin99,Marago2000} (see also Ref. \cite{PfauScissor} for the investigation of the scissor mode in dipolar droplets).

\section*{Open issues} 

The research activities on supersolidity are in continuous expansion. From a perspective point of view we envisage significant advances in the field concerning a series of relevant topics.

\begin{figure}[h]
\centering
\includegraphics[width=0.5\linewidth]{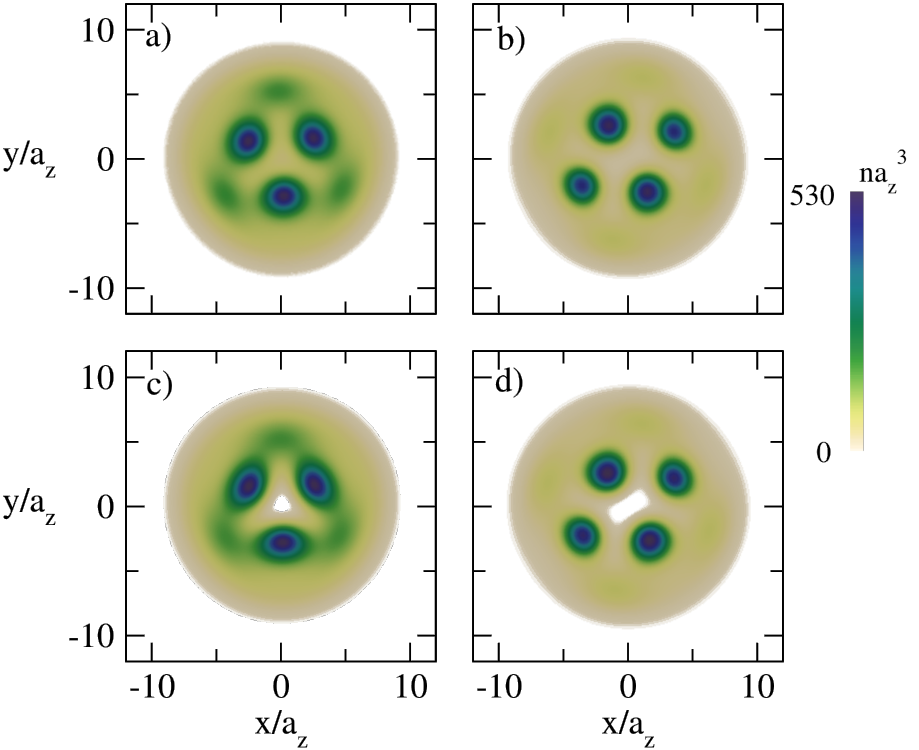}
\caption{Density plots of the vortical configurations in the case of a single-triangular cell structure (left) for
$\varepsilon_{dd}=1.334$ and in the case of a double-triangular cell (right) structure
obtained for $\varepsilon=1.351$. The harmonic oscillator length along the strongly confinrd direction $z$ is $a_z=0.87\mu m$ and a total number of $4\times 10^4$ $^{164}$Dy atoms has been considered. [Adapted from Ref.\cite{roccuzzo2}]}
\label{fig:vortices}
\end{figure}

\paragraph{Quantized vortices.} 
Quantized vortices are a key hallmark of superfluidity and superconductivity. 
They are topological defects of the order parameter, and therefore robust with respect to perturbations in the $U(1)$ broken symmetry phase. Quantized vortices have been observed  in type-II superconductors (the Abrikosov vortex lattice)~\cite{AbrikosovNL}, in liquid helium~\cite{Donnelly}, in exciton-polariton superfluids~\cite{Carusotto2013} and in ultra-cold atomic gases, where they have been realized both in Bose-Einstein condensates and in superfluid Fermi gases \cite{BecBook2016}.  
Very recently they have been observed also in the superfluid phase of a dipolar gas~\cite{VortexIBK}. Their experimental observation in the supersolid phase of a dipolar gas would be important as a further undisputed proof of superfluidity. Furthermore the presence of density modulation reduces the superfluid density of the system and the droplets can have their own dynamics. Theoretically quantum vortices in the supersolid phase have been studied~\cite{roccuzzo2,gallemi20,ancilotto21} -- as many other static and dynamic properties -- by using a density functional theory, also known as extended Gross-Pitaevskii equation, which includes LHY correction in the equaton of state of the system~\cite{Wachter2016}. A number of peculiar properties have been already pointed out~\cite{roccuzzo2,gallemi20}: (i) the vortex-core's size is of the same order as the modulation wavelength of the supersolid and present a deformed shape with respect to the usual circular one, as shown in Fig. \ref{fig:vortices} for typical experimental parameters; (ii)  the angular momentum per particle carried by a single vortex in the bulk is smaller than $\hslash$, due to the superfluid density reduction and the droplets carry part of the angular moment in a rigid body fashion; (iii) finally in a trap geometry the critical rotation  frequency for a vortex nucleation is significantly smaller than for a standard superfluid and is still given by the frequency of the superfluid quadrupole mode. 
Experimentally a possible route to generate vortices in the supersolid phase could be achieved by tilting a magnetic field  towards a cylindrically symmetric pancake-shaped trap and setting it into rotation. This method has recently proven successful for generating quantized vortices in the superfluid phase of a dipolar gas \cite{VortexIBK}.  

\paragraph{Dipolar gases in box potentials and edge supersolidity.}
Since the first realisation of an atomic BEC, ultra-cold gases have been trapped by means of electromagnetic fields, which give rise to approximated harmonic potentials. The sample inhomogeneity due to the trapping is often considered detrimental for the study of the thermodynamic properties. More recently, thanks to the use of digital mirror devices,  box-like potentials have been also realised. 
In this respect strongly rotonized  dipolar gases and, even more, dipolar supersolids show a rather peculiar behaviour. Actually in a box the gas tends to accumulate close to the border, which can be seen as an effect of a sharp potential in presence of a rotonized spectrum \cite{Lu2010-dipolarbox}. The atoms on the border self-arrange in interesting structures \cite{roccuzzoedge} as, e.g.,  one-dimensional edge supersolid states coexisting with a low-density bulk superfluid. In the case of very large density the bulk can sustain a supersolid phase.
In an infinite system the bulk lattice is expected to be triangular at typical densities -- as already experimentally shown even for small samples~\cite{2DSS-NatureIBK}-- or honeycomb for larger densities \cite{Pohl2019}. In the presence of the box potential the geometry of the lattice is instead determined by the shape of the box itself, due to the long-range nature of the inter-atomic potential~\cite{roccuzzoedge}.

\paragraph{ Nature of the phase transitions at zero and finite temperature.} The nature of the superfluid/supersolid and supersolid/crystal phase transition in dipolar gases and the role of dimensionality  are far from being fully explored, also due to the present experimental difficulties to realize  configurations containing a significant number of  droplets.  Very recent works have provided first insight on the role of dimensionality of the superfluid/supersolid transition  and the nature of the dimensional crossover controlled by the transverse confinement and the atom number \cite{blakie2020,biagioni2022}.    First theoretical and experimental papers have also shown that it is possible to reach the supersolid phase in 2D configurations using evaporative cooling techniques  \cite{Bland2022_2D}, paving the way for future investigations of supersolidity at finite temperature. In particular the role of  temperature in establishing the nature of the phase transitions as well as the formation of the density  modulations and the global phase coherence of the system are important issues to be further explored. Finally, the understanding of the nature of the supersolid/crystal phase transition in this systems  has been the object of first investigations \cite{sohmen2021,baena2023}.
In particular lowering the temperature one first form the droplets and eventually the coherence between the droplets is established. This behaviour confirm the naive intuition that thermal fluctuations can easily destroy the coherence when the inter-droplet density is very small. Experimentally such a study is complicated by the high value of the density of the droplets characterizing the crystal phase, which causes the instability of the system due to three-body recombinations. First hints of the  superfluid-Mott insulator nature of the transition ~\cite{Boninsegni_SS2012} have been experimentally obtained in Ref.\cite{JJA-Ferlaino}. In this work it has been shown that the main features resulting from a quench from the supersolid state to the indepedent droplet (i.e., insulating) phase are captured by a simple Josephson junction array model, well studied in the context of phase transition from a long-range ordered (coherent) phase to an insulating (incoherent) phase.

\paragraph{Superfluid density and propagation of sound.} The relation between the quenching of superfluidity and the velocity of sound in spatially modulated superfluids has recently attracted  experimental interest  in the field of ultracold atomic gases (see Refs.\cite{taoarxiv2023,Chauveauarxiv2023} for  recent studies on Bose-Einstein condensates subject to an external periodic potential). The question is even more intriguing when the density modulation is caused by the spontaneous breaking of translational invariance as happens in a supersolid. In this case an additional sound mode is predicted to occur at zero temperature and the relation between the superfluid density and the sound velocity of the two  Goldstone modes involves both the bulk and the layer compressibility modulus \cite{AndreevLifshitz,Saslow,Dorsey2010,Hofmann2021}. Experiments on the propagation of sound in supersolid dipolar gases are then expected to provide further light on their  crystal nature   and possibly give quantitative access to the quenching of the superfluid density.  The identification  of Josephson effects\cite{JJA-Ferlaino} in the propagation of sound in these density modulated systems is also expected to provide useful information on the role of the superfluid density \cite{zapata1998}.

\paragraph{Dipolar gases mixtures and supersolidity}

Similarly to the case of  atomic gases interacting only with short range potentials, it has been recently possible to obtain mixtures of  dipolar condensates \cite{trautmann2018,durastane2020,politi2022}. Such platforms open new perspectives concerning supersolidity. 
In general a two-component mixture presents a rich phase diagram and spectrum which is the consequence of the hybridization between the bare modes of the single constituents. For instance two coupled superfluids show two collisionless sound modes generated by the hybridization of the Goldstone modes due to the breaking of the $U(1)\times U(1)$ symmetry of the system. The hybridized modes are related to the in-phase (total density) and the  out-of-phase (longitudinal spin) dynamics. 
For a strongly dipolar system, where a  supersolid phase can occur, the situation is more complex. For instance the modulation could occur in both components in the same way -- due to the instability of the roton in the density sector --  or out-of-phase -- due to the instability of the roton in the spin sector \cite{Li2022,bland2022}. One of the main feature of the latter phase, in which the two component form alternating domains, is that it can be stabilized without quantum fluctuations, implying the possibility to realise larger systems with longer lifetimes. 
The theoretical analysis of such platforms is growing very fast and first experimental evidences of supersolid behaviour in such  systems is expected to become available soon. An important goal will be the study of the excitations and the role of the density and spin branches in the various supersolid phases.

\paragraph{Supersolidity of dipolar gases in two dimensions}  In two dimensions, and in the presence of a magnetic field orthogonal to the plane, the dipole interaction has a purely repulsive character. New scenarios can occur if one considers a slightly rotated magnetic field.  First theoretical calculations, based on Monte Carlo simulations in the presence of tilting  angles smaller than the critical value needed to generate attractive components in the dipole interactions, have revealed the occurrence of peculiar stripe configurations \cite{bombin2017}. The question of the stripe coherence is still debated \cite{cinti2019,BKT2d-dipolarstripes}. 
These stripes are formed only for high density configurations where three-body recombination plays a crucial role in destabilizing the system. In a  recent paper the emergence of stripes has been explored  increasing the tilting angle above the critical value \cite{staudinger2022}, favouring the conditions  for the formation of self-bound and striped phases. The density required to realize these stripes turns out to be significantly smaller with respect to the purely repulsive case, yielding more promising perspectives from the experimental point of view. 
A still unexplored question is whether the resulting stripes are coherently coupled and consequently whether these new configurations exhibit supersolidity.
It is worth mentioning that striped phases have already been experimentally obtained in a pancake geometry, but coherence has not yet been observed~\cite{Wenzel2017}.


\section*{Other Platforms} 
In recent years, in addition to the dipolar atomic gases discussed in this perspective, other  platforms have emerged, where supersolidity has already been observed, or which are expected  to exhibit relevant supersolid features.

\subsection*{Raman Induced Spin-orbit coupled BEC gases.} By applying Raman laser beams, which couple different spin states of a Bose-Einsten condensed gas, it is possible to generate a spin-orbit Hamiltonian which exhibits translational invariance in the proper spin rotated frame. For proper values of the Raman coupling the translational symmetry  is spontaneously broken \cite{Ho2011,YunLi2012} and the novel supersolid phase is characterized by full miscibility of two spin states and by density modulations which take the form of stripes  (for reviews see, e.g., Refs. \cite{Zhai2015,TrentoRevSOC2015}]. The phase transition to the miscible phase was first identified in the pioneering paper by the Spielman team \cite{LinSOC2011}, while the resulting modulations were first experimentally detected in 2016  using Bragg reflection \cite{Li2017} and recently imaged using matterwave lensing techniques [Leticia Tarruell, private communication]. Similarly to the dipolar case, also in spin-orbit coupled BEC gases the first order phase transition from the superfluid to the supersolid phase is accompanied by the softenting of the energy gap of the roton excitation. A peculiar feature of  the novel Goldstone mode exhibited by the the supersolid phase  is given  by its peculiar spin nature, which is the consequence of the hybridization between the density and the spin degrees of freedom  \cite{yunspectrum2013,Yunbo2017,Geier2021}. Spin is then expected to provide a natural tool to explore the crystal nature of the novel Goldstone modes occurring in the supersolid phase of spin-orbit coupled BEC's \cite{Geier2023}.

\subsection*{BEC gases confined in optical resonators} 
Another platform which has been shown to have a supersolid phase consists of an atomic BEC coupled to optical resonators. The second order phase transition to a supersolid state is driven by a superradiance instability, where the cavity field is coherently populated and determines in a selfconsistent way the modulation of the atomic cloud (see the recent review \cite{mivehvar2021} and reference therein). Two are the configurations which show supersolidity: 1) BEC's coupled to two non-collinear linear cavities, where the continuous spatial symmetry is related to the weight on the two coherent cavity fields \cite{ETH_SS1,ETH_SS2}; 2)  BEC's coupled to a ring cavity, where the contonuos symmetry is related to the phase of the modulation\cite{Zimmerman2020}. 

The spectrum of the system in the standard BEC and in the supersolid phase has been experimentally extracted, showing in particular the presence of a zero-energy mode in the supersolid phase and the closure of the Higgs mode at the transotion point \cite{ETH_SS1,ETH_SS2}.
Due to the fact that the period is determined by the cavity field, which induces a infinite range potential among the atoms, the  supersolid is in these platforms incompressible -- as any crystal with infinite long range interaction -- and therefore only the zero-momentum Goldstone mode is present, being all the
finite momenta (crystal) excitations gapped. Such shortcoming has been very recently removed by realising a BEC coupled to a confocal optical resonator~\cite{Guo2021}. Such a new platform exhibits lattice phonons and provides promising perspectives for supersolidity.

A transition to an independent cluster phase should also be present in these systems. Indeed in the more standard case of superradiance crystallization breaking only a $\mathbb{Z}_2$ symmetry (and not a continuous symmetry required for a system to be supersolid) a transition to the Mott phase has already been observed \cite{Hemmerich2015}. 


\subsection*{Supersolidity in moving atomic condensates}

We believe it is relevant to include in this Section a rather peculiar state of matter very much related to a supersolid, predicted by L. P. Pitaevskii in 1984 and which could be observed using ultra-cold gases. The main idea is that a system of bosons moving faster than the Landau critical velocity can form a periodically modulated superfluid \cite{Pitaevskii1984}. Pitaevskii considered the case of roton condensation in $^4$He -- more recently confirmed by means of density functional theory\cite{AncilottoLev2005}. The same physics can be observed in ultra-cold Bose gases due again to roton-like condensation (``levons'') \cite{levons2012} or cnoidal-wave formation \cite{KdV1895}. In many respects (superfluid density, excitation spectrum) these states can be seen as supersolid states\cite{martone2021}.

{\subsection*{Ultra-cold gases in optical lattices and $^4$He on substrate}}
Another platform where the search of a supersolid phase has been very intense consists of ultra-cold gases in optical lattices. The idea of extending the concept of supersolidity to a lattice model dates back to the work of Matsuda and Tsuneto in 1970 \cite{Matsuda1970}, where an analogy between a supersolid phase for hard-core bosons and the anti-ferromagnetic state of a spin model was already pointed out. Such a phase is sometimes named "lattice supersolid" to specify that the state does not break any contiouns symmetry but only a descrete symmetry (see, Ref. \cite{Boninsegni2012, kottmann2021} and reference therein). An experimental realisation of lattice supersolid breaking a $\mathbb{Z}_2$ symmetry has been realised again by means of a BEC in a single linear optical cavity \cite{dickeSS2010-esslinger,dicke-roton2012}.

Recent result on lattice supersolidity have also been recently explored in $^4$He on graphene layers \cite{balibarLayer,Choi2021,physLayerSS}.


\subsection*{The Larkin–Ovchinnikov state}
In fermionic spin unbalanced systems superfluidity (or superconductivity in solid state platform) can occur in different ways. Among them the Larkin-Ovchinikov (LO) phase in which the order parameter remains real, but oscillates in sign with a period inversely proportional to the Fermi momentum unbalance can be considered as an example of supersolidity.
The LO state indeed breaks simultaneously both the $U(1)$ and the translational symmetry, presenting two Goldstone mode \cite{Radzi2009}. Such a state  it has been  the object of systematic investigations in the context of superconductors in high magnetic field \cite{casalbuoni2004}. The LO state is particularly favoured in one-dimensional system \cite{yang2001} and, in the context of ultra-cold gases, first signatures of this phase in the context of atomic physics has been pointed out in ~\cite{orso2006,Hulet2010}. Many more experimental evidences of such a state have been reported in layered superconducting materials (see the recent perspective~\cite{pavani2022}) and in particular in layered organic superconductors \cite{FFLO-organic}.

{\subsection*{Crust of neutron stars}
In the context of nuclear physics another quantum phase which exhibits a supersolid behavior is the pasta phase in the inner crust of neutron stars \cite{chamel2008}.  In particular the question of the effects of the band structure on the neutron superfluid density has been recently addressed in \cite{WatPethick2017}, where   the crucial role of two body interactions in limiting the reduction of the superfluid density  has been pointed out, thereby renewing interest  for pulsar glitch models.}

\vspace{2cm}
\thispagestyle{empty}

\noindent \textbf{Key points:} 

\begin{itemize}
    \item Supersolidity is a subtle peculiar state, which combines superfluid and solid properties, predicted more than 50 years ago in the context of solid Helium-4.
    \item Ultra-Cold gases are clean and flexible platforms, where both the 2-body interaction and the single particle term can be engineer in order to obtain new phase of matter. 
    \item While for Heluim-4 a supersolid phase does not seem to exist, a few years ago a number of experimental groups reported on evidence of supersolid features in spin-orbit coupled gases, in BEC gases inside optical resonators and in dipolar gases.
    \item The dipolar gas platforms have been proven appropriate to map out a number of properties of the supersolid phase.
    
\end{itemize}

\noindent \textbf{Website Summary:} 

Supersolidity is an intriguing state of matter which combines superfluid and crystal features. Never observed before, atomic gases have recently been proven to exhibit typical supersolid properties like spontaneous density modulations combined with coherence effects and the occurrence of novel Goldstone modes.

\bibliography{bib}

\begin{thebibliography}{100}
\urlstyle{rm}
\expandafter\ifx\csname url\endcsname\relax
  \def\url#1{\texttt{#1}}\fi
\expandafter\ifx\csname urlprefix\endcsname\relax\def\urlprefix{URL }\fi
\expandafter\ifx\csname doiprefix\endcsname\relax\def\doiprefix{DOI: }\fi
\providecommand{\bibinfo}[2]{#2}
\providecommand{\eprint}[2][]{\url{#2}}

\bibitem{GrossAP1960}
\bibinfo{author}{Gross, E.~P.}
\newblock \bibinfo{journal}{\bibinfo{title}{Quantum theory of interacting
  bosons}}.
\newblock {\emph{\JournalTitle{Annals of Physics}}}
  \textbf{\bibinfo{volume}{9}}, \bibinfo{pages}{292--324},
  \doiprefix\url{https://doi.org/10.1016/0003-4916(60)90033-6}
  (\bibinfo{year}{1960}).

\bibitem{AndreevLifshitz}
\bibinfo{author}{Andreev, A.~F.} \& \bibinfo{author}{Lifshitz, I.~M.}
\newblock \bibinfo{journal}{\bibinfo{title}{Quantum theory of defects in
  crystals}}.
\newblock {\emph{\JournalTitle{Sov. Phys. JETP}}}
  \textbf{\bibinfo{volume}{29}}, \bibinfo{pages}{1107} (\bibinfo{year}{1969}).

\bibitem{chester1970}
\bibinfo{author}{Chester, G.~V.}
\newblock \bibinfo{journal}{\bibinfo{title}{Speculations on bose-einstein
  condensation and quantum crystals}}.
\newblock {\emph{\JournalTitle{Phys. Rev. A}}} \textbf{\bibinfo{volume}{2}},
  \bibinfo{pages}{256--258}, \doiprefix\url{10.1103/PhysRevA.2.256}
  (\bibinfo{year}{1970}).

\bibitem{Leggett1970}
\bibinfo{author}{Leggett, A.~J.}
\newblock \bibinfo{journal}{\bibinfo{title}{Can a solid be "superfluid"?}}
\newblock {\emph{\JournalTitle{Phys. Rev. Lett.}}}
  \textbf{\bibinfo{volume}{25}}, \bibinfo{pages}{1543--1546},
  \doiprefix\url{10.1103/PhysRevLett.25.1543} (\bibinfo{year}{1970}).

\bibitem{Chan2}
\bibinfo{author}{Kim, D.~Y.} \& \bibinfo{author}{Chan, M. H.~W.}
\newblock \bibinfo{journal}{\bibinfo{title}{Absence of supersolidity in solid
  helium in porous vycor glass}}.
\newblock {\emph{\JournalTitle{Phys. Rev. Lett.}}}
  \textbf{\bibinfo{volume}{109}}, \bibinfo{pages}{155301},
  \doiprefix\url{10.1103/PhysRevLett.109.155301} (\bibinfo{year}{2012}).

\bibitem{Balibar}
\bibinfo{author}{Balibar, S.}
\newblock \bibinfo{journal}{\bibinfo{title}{The enigma of supersolidity}}.
\newblock {\emph{\JournalTitle{Nature}}} \textbf{\bibinfo{volume}{464}},
  \bibinfo{pages}{176--182}, \doiprefix\url{10.1038/nature08913}
  (\bibinfo{year}{2010}).

\bibitem{nepo1971}
\bibinfo{author}{Kirzhnits, D.~A.} \& \bibinfo{author}{Nepomnyashchii, Y.~A.}
\newblock \bibinfo{journal}{\bibinfo{title}{Coherent crystallization of quantum
  liquid}}.
\newblock {\emph{\JournalTitle{Soviet Physics JETP}}}
  \textbf{\bibinfo{volume}{32}} (\bibinfo{year}{1971}).

\bibitem{Hydro-Rica2007}
\bibinfo{author}{Josserand, C.}, \bibinfo{author}{Pomeau, Y.} \&
  \bibinfo{author}{Rica, S.}
\newblock \bibinfo{journal}{\bibinfo{title}{Coexistence of ordinary elasticity
  and superfluidity in a model of a defect-free supersolid}}.
\newblock {\emph{\JournalTitle{Phys. Rev. Lett.}}}
  \textbf{\bibinfo{volume}{98}}, \bibinfo{pages}{195301},
  \doiprefix\url{10.1103/PhysRevLett.98.195301} (\bibinfo{year}{2007}).

\bibitem{Boninsegni2012}
\bibinfo{author}{Boninsegni, M.} \& \bibinfo{author}{Prokof'ev, N.~V.}
\newblock \bibinfo{journal}{\bibinfo{title}{Colloquium: Supersolids: What and
  where are they?}}
\newblock {\emph{\JournalTitle{Rev. Mod. Phys.}}}
  \textbf{\bibinfo{volume}{84}}, \bibinfo{pages}{759--776},
  \doiprefix\url{10.1103/RevModPhys.84.759} (\bibinfo{year}{2012}).

\bibitem{Li2017}
\bibinfo{author}{Li, J.-R.} \emph{et~al.}
\newblock \bibinfo{journal}{\bibinfo{title}{A stripe phase with supersolid
  properties in spin--orbit-coupled bose--einstein condensates}}.
\newblock {\emph{\JournalTitle{Nature}}} \textbf{\bibinfo{volume}{543}},
  \bibinfo{pages}{91--94}, \doiprefix\url{10.1038/nature21431}
  (\bibinfo{year}{2017}).

\bibitem{Leonard2017}
\bibinfo{author}{L{\'e}onard, J.}, \bibinfo{author}{Morales, A.},
  \bibinfo{author}{Zupancic, P.}, \bibinfo{author}{Esslinger, T.} \&
  \bibinfo{author}{Donner, T.}
\newblock \bibinfo{journal}{\bibinfo{title}{Supersolid formation in a quantum
  gas breaking a continuous translational symmetry}}.
\newblock {\emph{\JournalTitle{Nature}}} \textbf{\bibinfo{volume}{543}},
  \bibinfo{pages}{87--90}, \doiprefix\url{10.1038/nature21067}
  (\bibinfo{year}{2017}).

\bibitem{Pisa1}
\bibinfo{author}{Tanzi, L.} \emph{et~al.}
\newblock \bibinfo{journal}{\bibinfo{title}{Observation of a dipolar quantum
  gas with metastable supersolid properties}}.
\newblock {\emph{\JournalTitle{Phys. Rev. Lett.}}}
  \textbf{\bibinfo{volume}{122}}, \bibinfo{pages}{130405},
  \doiprefix\url{10.1103/PhysRevLett.122.130405} (\bibinfo{year}{2019}).

\bibitem{Stut1}
\bibinfo{author}{B\"ottcher, F.} \emph{et~al.}
\newblock \bibinfo{journal}{\bibinfo{title}{Transient supersolid properties in
  an array of dipolar quantum droplets}}.
\newblock {\emph{\JournalTitle{Phys. Rev. X}}} \textbf{\bibinfo{volume}{9}},
  \bibinfo{pages}{011051}, \doiprefix\url{10.1103/PhysRevX.9.011051}
  (\bibinfo{year}{2019}).

\bibitem{IBK1}
\bibinfo{author}{Chomaz, L.} \emph{et~al.}
\newblock \bibinfo{journal}{\bibinfo{title}{Long-lived and transient supersolid
  behaviors in dipolar quantum gases}}.
\newblock {\emph{\JournalTitle{Phys. Rev. X}}} \textbf{\bibinfo{volume}{9}},
  \bibinfo{pages}{021012}, \doiprefix\url{10.1103/PhysRevX.9.021012}
  (\bibinfo{year}{2019}).

\bibitem{Pomeau93}
\bibinfo{author}{Pomeau, Y.} \& \bibinfo{author}{Rica, S.}
\newblock \bibinfo{journal}{\bibinfo{title}{Model of superflow with rotons}}.
\newblock {\emph{\JournalTitle{Phys. Rev. Lett.}}}
  \textbf{\bibinfo{volume}{71}}, \bibinfo{pages}{247--250},
  \doiprefix\url{10.1103/PhysRevLett.71.247} (\bibinfo{year}{1993}).

\bibitem{Pomeau94}
\bibinfo{author}{Pomeau, Y.} \& \bibinfo{author}{Rica, S.}
\newblock \bibinfo{journal}{\bibinfo{title}{Dynamics of a model of
  supersolid}}.
\newblock {\emph{\JournalTitle{Phys. Rev. Lett.}}}
  \textbf{\bibinfo{volume}{72}}, \bibinfo{pages}{2426--2429},
  \doiprefix\url{10.1103/PhysRevLett.72.2426} (\bibinfo{year}{1994}).

\bibitem{Boninsegni_SS2012}
\bibinfo{author}{Saccani, S.}, \bibinfo{author}{Moroni, S.} \&
  \bibinfo{author}{Boninsegni, M.}
\newblock \bibinfo{journal}{\bibinfo{title}{Excitation spectrum of a
  supersolid}}.
\newblock {\emph{\JournalTitle{Phys. Rev. Lett.}}}
  \textbf{\bibinfo{volume}{108}}, \bibinfo{pages}{175301},
  \doiprefix\url{10.1103/PhysRevLett.108.175301} (\bibinfo{year}{2012}).

\bibitem{Chomaz_Review2023}
\bibinfo{author}{Chomaz, L.} \emph{et~al.}
\newblock \bibinfo{journal}{\bibinfo{title}{Dipolar physics: a review of
  experiments with magnetic quantum gases}}.
\newblock {\emph{\JournalTitle{Reports on Progress in Physics}}}
  \textbf{\bibinfo{volume}{86}}, \bibinfo{pages}{026401},
  \doiprefix\url{10.1088/1361-6633/aca814} (\bibinfo{year}{2022}).

\bibitem{SantosRoton}
\bibinfo{author}{Santos, L.}, \bibinfo{author}{Shlyapnikov, G.~V.} \&
  \bibinfo{author}{Lewenstein, M.}
\newblock \bibinfo{journal}{\bibinfo{title}{Roton-maxon spectrum and stability
  of trapped dipolar bose-einstein condensates}}.
\newblock {\emph{\JournalTitle{Phys. Rev. Lett.}}}
  \textbf{\bibinfo{volume}{90}}, \bibinfo{pages}{250403},
  \doiprefix\url{10.1103/PhysRevLett.90.250403} (\bibinfo{year}{2003}).

\bibitem{Odell2003}
\bibinfo{author}{O'Dell, D. H.~J.}, \bibinfo{author}{Giovanazzi, S.} \&
  \bibinfo{author}{Kurizki, G.}
\newblock \bibinfo{journal}{\bibinfo{title}{Rotons in gaseous bose-einstein
  condensates irradiated by a laser}}.
\newblock {\emph{\JournalTitle{Phys. Rev. Lett.}}}
  \textbf{\bibinfo{volume}{90}}, \bibinfo{pages}{110402},
  \doiprefix\url{10.1103/PhysRevLett.90.110402} (\bibinfo{year}{2003}).

\bibitem{Bisset2019}
\bibinfo{author}{Bisset, R.~N.}, \bibinfo{author}{Blakie, P.~B.} \&
  \bibinfo{author}{Stringari, S.}
\newblock \bibinfo{journal}{\bibinfo{title}{Static-response theory and the
  roton-maxon spectrum of a flattened dipolar bose-einstein condensate}}.
\newblock {\emph{\JournalTitle{Phys. Rev. A}}} \textbf{\bibinfo{volume}{100}},
  \bibinfo{pages}{013620}, \doiprefix\url{10.1103/PhysRevA.100.013620}
  (\bibinfo{year}{2019}).

\bibitem{celli1972}
\bibinfo{author}{Celli, V.} \& \bibinfo{author}{Ruvalds, J.}
\newblock \bibinfo{journal}{\bibinfo{title}{Theory of the liquid-solid phase
  transition in helium ii}}.
\newblock {\emph{\JournalTitle{Phys. Rev. Lett.}}}
  \textbf{\bibinfo{volume}{28}}, \bibinfo{pages}{539--542},
  \doiprefix\url{10.1103/PhysRevLett.28.539} (\bibinfo{year}{1972}).

\bibitem{balibar2006}
\bibinfo{author}{Balibar, S.}
\newblock \bibinfo{journal}{\bibinfo{title}{{Rotons, Superfluidity, and Helium
  Crystals}}}.
\newblock {\emph{\JournalTitle{AIP Conference Proceedings}}}
  \textbf{\bibinfo{volume}{850}}, \bibinfo{pages}{18--25},
  \doiprefix\url{10.1063/1.2354593} (\bibinfo{year}{2006}).
\newblock
  \eprint{https://pubs.aip.org/aip/acp/article-pdf/850/1/18/11577074/18\_1\_online.pdf}.

\bibitem{LHY1957}
\bibinfo{author}{Lee, T.~D.}, \bibinfo{author}{Huang, K.} \&
  \bibinfo{author}{Yang, C.~N.}
\newblock \bibinfo{journal}{\bibinfo{title}{Eigenvalues and eigenfunctions of a
  bose system of hard spheres and its low-temperature properties}}.
\newblock {\emph{\JournalTitle{Phys. Rev.}}} \textbf{\bibinfo{volume}{106}},
  \bibinfo{pages}{1135--1145}, \doiprefix\url{10.1103/PhysRev.106.1135}
  (\bibinfo{year}{1957}).

\bibitem{FischerLHY}
\bibinfo{author}{Sch\"utzhold, R.}, \bibinfo{author}{Uhlmann, M.},
  \bibinfo{author}{Xu, Y.} \& \bibinfo{author}{Fischer, U.~R.}
\newblock \bibinfo{journal}{\bibinfo{title}{Mean-field expansion in
  {Bose-Einstein} condensates with finite-range interactions}}.
\newblock {\emph{\JournalTitle{International Journal of Modern Physics B}}}
  \textbf{\bibinfo{volume}{20}}, \bibinfo{pages}{3555--3565},
  \doiprefix\url{10.1142/S0217979206035631} (\bibinfo{year}{2006}).

\bibitem{Pelster}
\bibinfo{author}{Lima, A. R.~P.} \& \bibinfo{author}{Pelster, A.}
\newblock \bibinfo{journal}{\bibinfo{title}{Beyond mean-field low-lying
  excitations of dipolar bose gases}}.
\newblock {\emph{\JournalTitle{Phys. Rev. A}}} \textbf{\bibinfo{volume}{86}},
  \bibinfo{pages}{063609}, \doiprefix\url{10.1103/PhysRevA.86.063609}
  (\bibinfo{year}{2012}).

\bibitem{PetrovDrop}
\bibinfo{author}{Petrov, D.~S.}
\newblock \bibinfo{journal}{\bibinfo{title}{Quantum mechanical stabilization of
  a collapsing bose-bose mixture}}.
\newblock {\emph{\JournalTitle{Phys. Rev. Lett.}}}
  \textbf{\bibinfo{volume}{115}}, \bibinfo{pages}{155302},
  \doiprefix\url{10.1103/PhysRevLett.115.155302} (\bibinfo{year}{2015}).

\bibitem{Wachter2016}
\bibinfo{author}{W\"achtler, F.} \& \bibinfo{author}{Santos, L.}
\newblock \bibinfo{journal}{\bibinfo{title}{Quantum filaments in dipolar
  bose-einstein condensates}}.
\newblock {\emph{\JournalTitle{Phys. Rev. A}}} \textbf{\bibinfo{volume}{93}},
  \bibinfo{pages}{061603}, \doiprefix\url{10.1103/PhysRevA.93.061603}
  (\bibinfo{year}{2016}).

\bibitem{BarbutDrop2016}
\bibinfo{author}{Ferrier-Barbut, I.}, \bibinfo{author}{Kadau, H.},
  \bibinfo{author}{Schmitt, M.}, \bibinfo{author}{Wenzel, M.} \&
  \bibinfo{author}{Pfau, T.}
\newblock \bibinfo{journal}{\bibinfo{title}{Observation of quantum droplets in
  a strongly dipolar bose gas}}.
\newblock {\emph{\JournalTitle{Phys. Rev. Lett.}}}
  \textbf{\bibinfo{volume}{116}}, \bibinfo{pages}{215301},
  \doiprefix\url{10.1103/PhysRevLett.116.215301} (\bibinfo{year}{2016}).

\bibitem{holger2016}
\bibinfo{author}{Kadau, H.} \emph{et~al.}
\newblock \bibinfo{journal}{\bibinfo{title}{Observing the rosensweig
  instability of a quantum ferrofluid}}.
\newblock {\emph{\JournalTitle{Nature}}} \textbf{\bibinfo{volume}{530}},
  \bibinfo{pages}{194--197}, \doiprefix\url{10.1038/nature16485}
  (\bibinfo{year}{2016}).

\bibitem{Schmitt2016}
\bibinfo{author}{Schmitt, M.}, \bibinfo{author}{Wenzel, M.},
  \bibinfo{author}{B{\"o}ttcher, F.}, \bibinfo{author}{Ferrier-Barbut, I.} \&
  \bibinfo{author}{Pfau, T.}
\newblock \bibinfo{journal}{\bibinfo{title}{Self-bound droplets of a dilute
  magnetic quantum liquid}}.
\newblock {\emph{\JournalTitle{Nature}}} \textbf{\bibinfo{volume}{539}},
  \bibinfo{pages}{259--262}, \doiprefix\url{10.1038/nature20126}
  (\bibinfo{year}{2016}).

\bibitem{ChomazDroplet2016}
\bibinfo{author}{Chomaz, L.} \emph{et~al.}
\newblock \bibinfo{journal}{\bibinfo{title}{Quantum-fluctuation-driven
  crossover from a dilute bose-einstein condensate to a macrodroplet in a
  dipolar quantum fluid}}.
\newblock {\emph{\JournalTitle{Phys. Rev. X}}} \textbf{\bibinfo{volume}{6}},
  \bibinfo{pages}{041039}, \doiprefix\url{10.1103/PhysRevX.6.041039}
  (\bibinfo{year}{2016}).

\bibitem{FattoriDrop}
\bibinfo{author}{Semeghini, G.} \emph{et~al.}
\newblock \bibinfo{journal}{\bibinfo{title}{Self-bound quantum droplets of
  atomic mixtures in free space}}.
\newblock {\emph{\JournalTitle{Phys. Rev. Lett.}}}
  \textbf{\bibinfo{volume}{120}}, \bibinfo{pages}{235301},
  \doiprefix\url{10.1103/PhysRevLett.120.235301} (\bibinfo{year}{2018}).

\bibitem{TarruellDrop}
\bibinfo{author}{Cabrera, C.~R.} \emph{et~al.}
\newblock \bibinfo{journal}{\bibinfo{title}{Quantum liquid droplets in a
  mixture of bose-einstein condensates}}.
\newblock {\emph{\JournalTitle{Science}}} \textbf{\bibinfo{volume}{359}},
  \bibinfo{pages}{301--304}, \doiprefix\url{10.1126/science.aao5686}
  (\bibinfo{year}{2018}).
\newblock
  \eprint{https://science.sciencemag.org/content/359/6373/301.full.pdf}.

\bibitem{TarruellDrop2}
\bibinfo{author}{Cheiney, P.} \emph{et~al.}
\newblock \bibinfo{journal}{\bibinfo{title}{Bright soliton to quantum droplet
  transition in a mixture of bose-einstein condensates}}.
\newblock {\emph{\JournalTitle{Phys. Rev. Lett.}}}
  \textbf{\bibinfo{volume}{120}}, \bibinfo{pages}{135301},
  \doiprefix\url{10.1103/PhysRevLett.120.135301} (\bibinfo{year}{2018}).

\bibitem{FattoriColl}
\bibinfo{author}{Ferioli, G.} \emph{et~al.}
\newblock \bibinfo{journal}{\bibinfo{title}{Collisions of self-bound quantum
  droplets}}.
\newblock {\emph{\JournalTitle{Phys. Rev. Lett.}}}
  \textbf{\bibinfo{volume}{122}}, \bibinfo{pages}{090401},
  \doiprefix\url{10.1103/PhysRevLett.122.090401} (\bibinfo{year}{2019}).

\bibitem{IBKrotons}
\bibinfo{author}{Chomaz, L.} \emph{et~al.}
\newblock \bibinfo{journal}{\bibinfo{title}{{Observation of roton mode
  population in a dipolar quantum gas}}}.
\newblock {\emph{\JournalTitle{Nature Physics}}} \textbf{\bibinfo{volume}{14}},
  \bibinfo{pages}{442--446}, \doiprefix\url{10.1038/s41567-018-0054-7}
  (\bibinfo{year}{2018}).

\bibitem{petter-roton2019}
\bibinfo{author}{Petter, D.} \emph{et~al.}
\newblock \bibinfo{journal}{\bibinfo{title}{Probing the roton excitation
  spectrum of a stable dipolar bose gas}}.
\newblock {\emph{\JournalTitle{Phys. Rev. Lett.}}}
  \textbf{\bibinfo{volume}{122}}, \bibinfo{pages}{183401},
  \doiprefix\url{10.1103/PhysRevLett.122.183401} (\bibinfo{year}{2019}).

\bibitem{Roton2DPfau2021}
\bibinfo{author}{Schmidt, J.-N.} \emph{et~al.}
\newblock \bibinfo{journal}{\bibinfo{title}{Roton excitations in an oblate
  dipolar quantum gas}}.
\newblock {\emph{\JournalTitle{Phys. Rev. Lett.}}}
  \textbf{\bibinfo{volume}{126}}, \bibinfo{pages}{193002},
  \doiprefix\url{10.1103/PhysRevLett.126.193002} (\bibinfo{year}{2021}).

\bibitem{Roccuzzo1}
\bibinfo{author}{Roccuzzo, S.~M.} \& \bibinfo{author}{Ancilotto, F.}
\newblock \bibinfo{journal}{\bibinfo{title}{Supersolid behavior of a dipolar
  bose-einstein condensate confined in a tube}}.
\newblock {\emph{\JournalTitle{Phys. Rev. A}}} \textbf{\bibinfo{volume}{99}},
  \bibinfo{pages}{041601}, \doiprefix\url{10.1103/PhysRevA.99.041601}
  (\bibinfo{year}{2019}).

\bibitem{Norcia2021_2D}
\bibinfo{author}{Norcia, M.~A.} \emph{et~al.}
\newblock \bibinfo{journal}{\bibinfo{title}{Two-dimensional supersolidity in a
  dipolar quantum gas}}.
\newblock {\emph{\JournalTitle{Nature}}} \textbf{\bibinfo{volume}{596}},
  \bibinfo{pages}{357--361}, \doiprefix\url{10.1038/s41586-021-03725-7}
  (\bibinfo{year}{2021}).

\bibitem{Bland2022_2D}
\bibinfo{author}{Bland, T.} \emph{et~al.}
\newblock \bibinfo{journal}{\bibinfo{title}{Two-dimensional supersolid
  formation in dipolar condensates}}.
\newblock {\emph{\JournalTitle{Phys. Rev. Lett.}}}
  \textbf{\bibinfo{volume}{128}}, \bibinfo{pages}{195302},
  \doiprefix\url{10.1103/PhysRevLett.128.195302} (\bibinfo{year}{2022}).

\bibitem{Macri_SS}
\bibinfo{author}{Macr\`{\i}, T.}, \bibinfo{author}{Maucher, F.},
  \bibinfo{author}{Cinti, F.} \& \bibinfo{author}{Pohl, T.}
\newblock \bibinfo{journal}{\bibinfo{title}{Elementary excitations of ultracold
  soft-core bosons across the superfluid-supersolid phase transition}}.
\newblock {\emph{\JournalTitle{Phys. Rev. A}}} \textbf{\bibinfo{volume}{87}},
  \bibinfo{pages}{061602}, \doiprefix\url{10.1103/PhysRevA.87.061602}
  (\bibinfo{year}{2013}).

\bibitem{Ancilotto2013}
\bibinfo{author}{Ancilotto, F.}, \bibinfo{author}{Rossi, M.} \&
  \bibinfo{author}{Toigo, F.}
\newblock \bibinfo{journal}{\bibinfo{title}{Supersolid structure and excitation
  spectrum of soft-core bosons in three dimensions}}.
\newblock {\emph{\JournalTitle{Phys. Rev. A}}} \textbf{\bibinfo{volume}{88}},
  \bibinfo{pages}{033618}, \doiprefix\url{10.1103/PhysRevA.88.033618}
  (\bibinfo{year}{2013}).

\bibitem{Pisa2}
\bibinfo{author}{Tanzi, L.} \emph{et~al.}
\newblock \bibinfo{journal}{\bibinfo{title}{{Supersolid symmetry breaking from
  compressional oscillations in a dipolar quantum gas}}}.
\newblock {\emph{\JournalTitle{Nature}}} \textbf{\bibinfo{volume}{574}},
  \bibinfo{pages}{382}, \doiprefix\url{10.1038/s41586-019-1568-6}
  (\bibinfo{year}{2019}).

\bibitem{Stut2}
\bibinfo{author}{Guo, M.} \emph{et~al.}
\newblock \bibinfo{journal}{\bibinfo{title}{{The low-energy Goldstone mode in a
  trapped dipolar supersolid}}}.
\newblock {\emph{\JournalTitle{Nature}}} \textbf{\bibinfo{volume}{574}},
  \bibinfo{pages}{386}, \doiprefix\url{10.1038/s41586-019-1569-5}
  (\bibinfo{year}{2019}).

\bibitem{NatNews2019}
\bibinfo{author}{Mossman, S.~M.}
\newblock \bibinfo{journal}{\bibinfo{title}{{Sounds of a supersolid detected in
  dipolar atomic gases for the first time}}}.
\newblock {\emph{\JournalTitle{Nature}}} \textbf{\bibinfo{volume}{574}},
  \bibinfo{pages}{382}, \doiprefix\url{10.1038/d41586-019-03045-x}
  (\bibinfo{year}{2019}).

\bibitem{Ibk2}
\bibinfo{author}{Natale, G.} \emph{et~al.}
\newblock \bibinfo{journal}{\bibinfo{title}{Excitation spectrum of a trapped
  dipolar supersolid and its experimental evidence}}.
\newblock {\emph{\JournalTitle{Phys. Rev. Lett.}}}
  \textbf{\bibinfo{volume}{123}}, \bibinfo{pages}{050402},
  \doiprefix\url{10.1103/PhysRevLett.123.050402} (\bibinfo{year}{2019}).

\bibitem{2DSS-NatureIBK}
\bibinfo{author}{Norcia, M.~A.} \emph{et~al.}
\newblock \bibinfo{journal}{\bibinfo{title}{Two-dimensional supersolidity in a
  dipolar quantum gas}}.
\newblock {\emph{\JournalTitle{Nature}}} \textbf{\bibinfo{volume}{596}},
  \bibinfo{pages}{357--361}, \doiprefix\url{10.1038/s41586-021-03725-7}
  (\bibinfo{year}{2021}).

\bibitem{biagioni2022}
\bibinfo{author}{Biagioni, G.} \emph{et~al.}
\newblock \bibinfo{journal}{\bibinfo{title}{Dimensional crossover in the
  superfluid-supersolid quantum phase transition}}.
\newblock {\emph{\JournalTitle{Phys. Rev. X}}} \textbf{\bibinfo{volume}{12}},
  \bibinfo{pages}{021019}, \doiprefix\url{10.1103/PhysRevX.12.021019}
  (\bibinfo{year}{2022}).

\bibitem{Norcia2021_scissors}
\bibinfo{author}{Norcia, M.~A.} \emph{et~al.}
\newblock \bibinfo{journal}{\bibinfo{title}{Can angular oscillations probe
  superfluidity in dipolar supersolids?}}
\newblock {\emph{\JournalTitle{Phys. Rev. Lett.}}}
  \textbf{\bibinfo{volume}{129}}, \bibinfo{pages}{040403},
  \doiprefix\url{10.1103/PhysRevLett.129.040403} (\bibinfo{year}{2022}).

\bibitem{Roccuzzo2021_inertia}
\bibinfo{author}{Roccuzzo, S.~M.}, \bibinfo{author}{Recati, A.} \&
  \bibinfo{author}{Stringari, S.}
\newblock \bibinfo{journal}{\bibinfo{title}{Moment of inertia and dynamical
  rotational response of a supersolid dipolar gas}}.
\newblock {\emph{\JournalTitle{Phys. Rev. A}}} \textbf{\bibinfo{volume}{105}},
  \bibinfo{pages}{023316}, \doiprefix\url{10.1103/PhysRevA.105.023316}
  (\bibinfo{year}{2022}).

\bibitem{roccuzzo2}
\bibinfo{author}{Roccuzzo, S.~M.}, \bibinfo{author}{Gallem\'{\i}, A.},
  \bibinfo{author}{Recati, A.} \& \bibinfo{author}{Stringari, S.}
\newblock \bibinfo{journal}{\bibinfo{title}{Rotating a supersolid dipolar
  gas}}.
\newblock {\emph{\JournalTitle{Phys. Rev. Lett.}}}
  \textbf{\bibinfo{volume}{124}}, \bibinfo{pages}{045702},
  \doiprefix\url{10.1103/PhysRevLett.124.045702} (\bibinfo{year}{2020}).

\bibitem{ScissorPisaSS}
\bibinfo{author}{Tanzi, L.} \emph{et~al.}
\newblock \bibinfo{journal}{\bibinfo{title}{Evidence of superfluidity in a
  dipolar supersolid from nonclassical rotational inertia}}.
\newblock {\emph{\JournalTitle{Science}}} \textbf{\bibinfo{volume}{371}},
  \bibinfo{pages}{1162--1165}, \doiprefix\url{10.1126/science.aba4309}
  (\bibinfo{year}{2021}).
\newblock \eprint{https://www.science.org/doi/pdf/10.1126/science.aba4309}.

\bibitem{Odelin99}
\bibinfo{author}{Gu\'ery-Odelin, D.} \& \bibinfo{author}{Stringari, S.}
\newblock \bibinfo{journal}{\bibinfo{title}{Scissors mode and superfluidity of
  a trapped bose-einstein condensed gas}}.
\newblock {\emph{\JournalTitle{Phys. Rev. Lett.}}}
  \textbf{\bibinfo{volume}{83}}, \bibinfo{pages}{4452--4455},
  \doiprefix\url{10.1103/PhysRevLett.83.4452} (\bibinfo{year}{1999}).

\bibitem{Marago2000}
\bibinfo{author}{Marag\`o, O.~M.} \emph{et~al.}
\newblock \bibinfo{journal}{\bibinfo{title}{Observation of the scissors mode
  and evidence for superfluidity of a trapped bose-einstein condensed gas}}.
\newblock {\emph{\JournalTitle{Phys. Rev. Lett.}}}
  \textbf{\bibinfo{volume}{84}}, \bibinfo{pages}{2056--2059},
  \doiprefix\url{10.1103/PhysRevLett.84.2056} (\bibinfo{year}{2000}).

\bibitem{PfauScissor}
\bibinfo{author}{Ferrier-Barbut, I.} \emph{et~al.}
\newblock \bibinfo{journal}{\bibinfo{title}{Scissors mode of dipolar quantum
  droplets of dysprosium atoms}}.
\newblock {\emph{\JournalTitle{Phys. Rev. Lett.}}}
  \textbf{\bibinfo{volume}{120}}, \bibinfo{pages}{160402},
  \doiprefix\url{10.1103/PhysRevLett.120.160402} (\bibinfo{year}{2018}).

\bibitem{AbrikosovNL}
\bibinfo{author}{Abrikosov, A.~A.}
\newblock \bibinfo{journal}{\bibinfo{title}{Nobel lecture: Type-ii
  superconductors and the vortex lattice}}.
\newblock {\emph{\JournalTitle{Rev. Mod. Phys.}}}
  \textbf{\bibinfo{volume}{76}}, \bibinfo{pages}{975--979},
  \doiprefix\url{10.1103/RevModPhys.76.975} (\bibinfo{year}{2004}).

\bibitem{Donnelly}
\bibinfo{author}{Donnelly, J.~R.}
\newblock \emph{\bibinfo{title}{Quantized Vortices in Heluim II}}
  (\bibinfo{publisher}{Cambridge University Press}, \bibinfo{year}{1991}).

\bibitem{Carusotto2013}
\bibinfo{author}{Carusotto, I.} \& \bibinfo{author}{Ciuti, C.}
\newblock \bibinfo{journal}{\bibinfo{title}{Quantum fluids of light}}.
\newblock {\emph{\JournalTitle{Rev. Mod. Phys.}}}
  \textbf{\bibinfo{volume}{85}}, \bibinfo{pages}{299--366},
  \doiprefix\url{10.1103/RevModPhys.85.299} (\bibinfo{year}{2013}).

\bibitem{BecBook2016}
\bibinfo{author}{Pitaevskii, L.} \& \bibinfo{author}{Stringari, S.}
\newblock \emph{\bibinfo{title}{Bose-Einstein condensation and superfluidity}}
  (\bibinfo{publisher}{Oxford University Press}, \bibinfo{year}{2016}).

\bibitem{VortexIBK}
\bibinfo{author}{Klaus, L.} \emph{et~al.}
\newblock \bibinfo{journal}{\bibinfo{title}{Observation of vortices and vortex
  stripes in a dipolar condensate}}.
\newblock {\emph{\JournalTitle{Nature Physics}}} \textbf{\bibinfo{volume}{18}},
  \bibinfo{pages}{1453--1458}, \doiprefix\url{10.1038/s41567-022-01793-8}
  (\bibinfo{year}{2022}).

\bibitem{gallemi20}
\bibinfo{author}{Gallem\'{\i}, A.}, \bibinfo{author}{Roccuzzo, S.~M.},
  \bibinfo{author}{Stringari, S.} \& \bibinfo{author}{Recati, A.}
\newblock \bibinfo{journal}{\bibinfo{title}{Quantized vortices in dipolar
  supersolid bose-einstein-condensed gases}}.
\newblock {\emph{\JournalTitle{Phys. Rev. A}}} \textbf{\bibinfo{volume}{102}},
  \bibinfo{pages}{023322}, \doiprefix\url{10.1103/PhysRevA.102.023322}
  (\bibinfo{year}{2020}).

\bibitem{ancilotto21}
\bibinfo{author}{Ancilotto, F.}, \bibinfo{author}{Barranco, M.},
  \bibinfo{author}{Pi, M.} \& \bibinfo{author}{Reatto, L.}
\newblock \bibinfo{journal}{\bibinfo{title}{Vortex properties in the extended
  supersolid phase of dipolar bose-einstein condensates}}.
\newblock {\emph{\JournalTitle{Phys. Rev. A}}} \textbf{\bibinfo{volume}{103}},
  \bibinfo{pages}{033314}, \doiprefix\url{10.1103/PhysRevA.103.033314}
  (\bibinfo{year}{2021}).

\bibitem{Lu2010-dipolarbox}
\bibinfo{author}{Lu, H.-Y.} \emph{et~al.}
\newblock \bibinfo{journal}{\bibinfo{title}{Spatial density oscillations in
  trapped dipolar condensates}}.
\newblock {\emph{\JournalTitle{Phys. Rev. A}}} \textbf{\bibinfo{volume}{82}},
  \bibinfo{pages}{023622}, \doiprefix\url{10.1103/PhysRevA.82.023622}
  (\bibinfo{year}{2010}).

\bibitem{roccuzzoedge}
\bibinfo{author}{Roccuzzo, S.~M.}, \bibinfo{author}{Stringari, S.} \&
  \bibinfo{author}{Recati, A.}
\newblock \bibinfo{journal}{\bibinfo{title}{Supersolid edge and bulk phases of
  a dipolar quantum gas in a box}}.
\newblock {\emph{\JournalTitle{Phys. Rev. Res.}}} \textbf{\bibinfo{volume}{4}},
  \bibinfo{pages}{013086}, \doiprefix\url{10.1103/PhysRevResearch.4.013086}
  (\bibinfo{year}{2022}).

\bibitem{Pohl2019}
\bibinfo{author}{Zhang, Y.-C.}, \bibinfo{author}{Maucher, F.} \&
  \bibinfo{author}{Pohl, T.}
\newblock \bibinfo{journal}{\bibinfo{title}{Supersolidity around a critical
  point in dipolar bose-einstein condensates}}.
\newblock {\emph{\JournalTitle{Phys. Rev. Lett.}}}
  \textbf{\bibinfo{volume}{123}}, \bibinfo{pages}{015301},
  \doiprefix\url{10.1103/PhysRevLett.123.015301} (\bibinfo{year}{2019}).

\bibitem{blakie2020}
\bibinfo{author}{Blakie, P.~B.}, \bibinfo{author}{Baillie, D.},
  \bibinfo{author}{Chomaz, L.} \& \bibinfo{author}{Ferlaino, F.}
\newblock \bibinfo{journal}{\bibinfo{title}{Supersolidity in an elongated
  dipolar condensate}}.
\newblock {\emph{\JournalTitle{Phys. Rev. Res.}}} \textbf{\bibinfo{volume}{2}},
  \bibinfo{pages}{043318}, \doiprefix\url{10.1103/PhysRevResearch.2.043318}
  (\bibinfo{year}{2020}).

\bibitem{sohmen2021}
\bibinfo{author}{Sohmen, M.} \emph{et~al.}
\newblock \bibinfo{journal}{\bibinfo{title}{Birth, life, and death of a dipolar
  supersolid}}.
\newblock {\emph{\JournalTitle{Phys. Rev. Lett.}}}
  \textbf{\bibinfo{volume}{126}}, \bibinfo{pages}{233401},
  \doiprefix\url{10.1103/PhysRevLett.126.233401} (\bibinfo{year}{2021}).

\bibitem{baena2023}
\bibinfo{author}{S{\'a}nchez-Baena, J.}, \bibinfo{author}{Politi, C.},
  \bibinfo{author}{Maucher, F.}, \bibinfo{author}{Ferlaino, F.} \&
  \bibinfo{author}{Pohl, T.}
\newblock \bibinfo{journal}{\bibinfo{title}{Heating a dipolar quantum fluid
  into a solid}}.
\newblock {\emph{\JournalTitle{Nature Communications}}}
  \textbf{\bibinfo{volume}{14}}, \bibinfo{pages}{1868},
  \doiprefix\url{10.1038/s41467-023-37207-3} (\bibinfo{year}{2023}).

\bibitem{JJA-Ferlaino}
\bibinfo{author}{Ilzh{\"o}fer, P.} \emph{et~al.}
\newblock \bibinfo{journal}{\bibinfo{title}{Phase coherence in
  out-of-equilibrium supersolid states of ultracold dipolar atoms}}.
\newblock {\emph{\JournalTitle{Nature Physics}}} \textbf{\bibinfo{volume}{17}},
  \bibinfo{pages}{356--361}, \doiprefix\url{10.1038/s41567-020-01100-3}
  (\bibinfo{year}{2021}).

\bibitem{taoarxiv2023}
\bibinfo{author}{Tao, J.}, \bibinfo{author}{Zhao, M.} \&
  \bibinfo{author}{Spielman, I.}
\newblock \bibinfo{title}{Observation of anisotropic superfluid density in an
  artificial crystal}, \doiprefix\url{10.48550/ARXIV.2301.01258}
  (\bibinfo{year}{2023}).

\bibitem{Chauveauarxiv2023}
\bibinfo{author}{Chauveau, G.} \emph{et~al.}
\newblock \bibinfo{title}{Superfluid fraction in an interacting spatially
  modulated bose-einstein condensate},
  \doiprefix\url{10.48550/ARXIV.2302.01776} (\bibinfo{year}{2023}).

\bibitem{Saslow}
\bibinfo{author}{Saslow, W.~M.}
\newblock \bibinfo{journal}{\bibinfo{title}{Microscopic and hydrodynamic theory
  of superfluidity in periodic solids}}.
\newblock {\emph{\JournalTitle{Phys. Rev. B}}} \textbf{\bibinfo{volume}{15}},
  \bibinfo{pages}{173--186}, \doiprefix\url{10.1103/PhysRevB.15.173}
  (\bibinfo{year}{1977}).

\bibitem{Dorsey2010}
\bibinfo{author}{Yoo, C.-D.} \& \bibinfo{author}{Dorsey, A.~T.}
\newblock \bibinfo{journal}{\bibinfo{title}{Hydrodynamic theory of supersolids:
  Variational principle, effective lagrangian, and density-density correlation
  function}}.
\newblock {\emph{\JournalTitle{Phys. Rev. B}}} \textbf{\bibinfo{volume}{81}},
  \bibinfo{pages}{134518}, \doiprefix\url{10.1103/PhysRevB.81.134518}
  (\bibinfo{year}{2010}).

\bibitem{Hofmann2021}
\bibinfo{author}{Hofmann, J.} \& \bibinfo{author}{Zwerger, W.}
\newblock \bibinfo{journal}{\bibinfo{title}{Hydrodynamics of a superfluid
  smectic}}.
\newblock {\emph{\JournalTitle{J. Stat. Mech.}}} \bibinfo{pages}{033104},
  \doiprefix\url{10.1088/1742-5468/abe598} (\bibinfo{year}{2021}).

\bibitem{zapata1998}
\bibinfo{author}{Zapata, I.}, \bibinfo{author}{Sols, F.} \&
  \bibinfo{author}{Leggett, A.~J.}
\newblock \bibinfo{journal}{\bibinfo{title}{Josephson effect between trapped
  bose-einstein condensates}}.
\newblock {\emph{\JournalTitle{Phys. Rev. A}}} \textbf{\bibinfo{volume}{57}},
  \bibinfo{pages}{R28--R31}, \doiprefix\url{10.1103/PhysRevA.57.R28}
  (\bibinfo{year}{1998}).

\bibitem{trautmann2018}
\bibinfo{author}{Trautmann, A.} \emph{et~al.}
\newblock \bibinfo{journal}{\bibinfo{title}{Dipolar quantum mixtures of erbium
  and dysprosium atoms}}.
\newblock {\emph{\JournalTitle{Phys. Rev. Lett.}}}
  \textbf{\bibinfo{volume}{121}}, \bibinfo{pages}{213601},
  \doiprefix\url{10.1103/PhysRevLett.121.213601} (\bibinfo{year}{2018}).

\bibitem{durastane2020}
\bibinfo{author}{Durastante, G.} \emph{et~al.}
\newblock \bibinfo{journal}{\bibinfo{title}{Feshbach resonances in an
  erbium-dysprosium dipolar mixture}}.
\newblock {\emph{\JournalTitle{Phys. Rev. A}}} \textbf{\bibinfo{volume}{102}},
  \bibinfo{pages}{033330}, \doiprefix\url{10.1103/PhysRevA.102.033330}
  (\bibinfo{year}{2020}).

\bibitem{politi2022}
\bibinfo{author}{Politi, C.} \emph{et~al.}
\newblock \bibinfo{journal}{\bibinfo{title}{Interspecies interactions in an
  ultracold dipolar mixture}}.
\newblock {\emph{\JournalTitle{Phys. Rev. A}}} \textbf{\bibinfo{volume}{105}},
  \bibinfo{pages}{023304}, \doiprefix\url{10.1103/PhysRevA.105.023304}
  (\bibinfo{year}{2022}).

\bibitem{Li2022}
\bibinfo{author}{Li, S.}, \bibinfo{author}{Le, U.~N.} \&
  \bibinfo{author}{Saito, H.}
\newblock \bibinfo{journal}{\bibinfo{title}{Long-lifetime supersolid in a
  two-component dipolar bose-einstein condensate}}.
\newblock {\emph{\JournalTitle{Phys. Rev. A}}} \textbf{\bibinfo{volume}{105}},
  \bibinfo{pages}{L061302}, \doiprefix\url{10.1103/PhysRevA.105.L061302}
  (\bibinfo{year}{2022}).

\bibitem{bland2022}
\bibinfo{author}{Bland, T.} \emph{et~al.}
\newblock \bibinfo{journal}{\bibinfo{title}{Alternating-domain supersolids in
  binary dipolar condensates}}.
\newblock {\emph{\JournalTitle{Phys. Rev. A}}} \textbf{\bibinfo{volume}{106}},
  \bibinfo{pages}{053322}, \doiprefix\url{10.1103/PhysRevA.106.053322}
  (\bibinfo{year}{2022}).

\bibitem{bombin2017}
\bibinfo{author}{Bombin, R.}, \bibinfo{author}{Boronat, J.} \&
  \bibinfo{author}{Mazzanti, F.}
\newblock \bibinfo{journal}{\bibinfo{title}{Dipolar bose supersolid stripes}}.
\newblock {\emph{\JournalTitle{Phys. Rev. Lett.}}}
  \textbf{\bibinfo{volume}{119}}, \bibinfo{pages}{250402},
  \doiprefix\url{10.1103/PhysRevLett.119.250402} (\bibinfo{year}{2017}).

\bibitem{cinti2019}
\bibinfo{author}{Cinti, F.} \& \bibinfo{author}{Boninsegni, M.}
\newblock \bibinfo{journal}{\bibinfo{title}{Absence of superfluidity in 2d
  dipolar bose striped crystals}}.
\newblock {\emph{\JournalTitle{Journal of Low Temperature Physics}}}
  \textbf{\bibinfo{volume}{196}}, \bibinfo{pages}{413--422},
  \doiprefix\url{10.1007/s10909-019-02209-3} (\bibinfo{year}{2019}).

\bibitem{BKT2d-dipolarstripes}
\bibinfo{author}{Bomb\'{\i}n, R.}, \bibinfo{author}{Mazzanti, F.} \&
  \bibinfo{author}{Boronat, J.}
\newblock \bibinfo{journal}{\bibinfo{title}{{Berezinskii-Kosterlitz-Thouless}
  transition in two-dimensional dipolar stripes}}.
\newblock {\emph{\JournalTitle{Phys. Rev. A}}} \textbf{\bibinfo{volume}{100}},
  \bibinfo{pages}{063614}, \doiprefix\url{10.1103/PhysRevA.100.063614}
  (\bibinfo{year}{2019}).

\bibitem{staudinger2022}
\bibinfo{author}{Staudinger, C.}, \bibinfo{author}{Hufnagl, D.},
  \bibinfo{author}{Mazzanti, F.} \& \bibinfo{author}{Zillich, R.~E.}
\newblock \bibinfo{title}{Striped ultradilute liquid of dipolar bosons in two
  dimensions}, \doiprefix\url{10.48550/ARXIV.2208.05028}
  (\bibinfo{year}{2022}).

\bibitem{Wenzel2017}
\bibinfo{author}{Wenzel, M.}, \bibinfo{author}{B\"ottcher, F.},
  \bibinfo{author}{Langen, T.}, \bibinfo{author}{Ferrier-Barbut, I.} \&
  \bibinfo{author}{Pfau, T.}
\newblock \bibinfo{journal}{\bibinfo{title}{Striped states in a many-body
  system of tilted dipoles}}.
\newblock {\emph{\JournalTitle{Phys. Rev. A}}} \textbf{\bibinfo{volume}{96}},
  \bibinfo{pages}{053630}, \doiprefix\url{10.1103/PhysRevA.96.053630}
  (\bibinfo{year}{2017}).

\bibitem{Ho2011}
\bibinfo{author}{Ho, T.-L.} \& \bibinfo{author}{Zhang, S.}
\newblock \bibinfo{journal}{\bibinfo{title}{Bose-einstein condensates with
  spin-orbit interaction}}.
\newblock {\emph{\JournalTitle{Phys. Rev. Lett.}}}
  \textbf{\bibinfo{volume}{107}}, \bibinfo{pages}{150403},
  \doiprefix\url{10.1103/PhysRevLett.107.150403} (\bibinfo{year}{2011}).

\bibitem{YunLi2012}
\bibinfo{author}{Li, Y.}, \bibinfo{author}{Pitaevskii, L.~P.} \&
  \bibinfo{author}{Stringari, S.}
\newblock \bibinfo{journal}{\bibinfo{title}{Quantum tricriticality and phase
  transitions in spin-orbit coupled bose-einstein condensates}}.
\newblock {\emph{\JournalTitle{Phys. Rev. Lett.}}}
  \textbf{\bibinfo{volume}{108}}, \bibinfo{pages}{225301},
  \doiprefix\url{10.1103/PhysRevLett.108.225301} (\bibinfo{year}{2012}).

\bibitem{Zhai2015}
\bibinfo{author}{Zhai, H.}
\newblock \bibinfo{journal}{\bibinfo{title}{Degenerate quantum gases with
  spin--orbit coupling: a review}}.
\newblock {\emph{\JournalTitle{Reports on Progress in Physics}}}
  \textbf{\bibinfo{volume}{78}}, \bibinfo{pages}{026001},
  \doiprefix\url{10.1088/0034-4885/78/2/026001} (\bibinfo{year}{2015}).

\bibitem{TrentoRevSOC2015}
\bibinfo{author}{Li, Y.}, \bibinfo{author}{Martone, G.~I.} \&
  \bibinfo{author}{Stringari, S.}
\newblock \emph{\bibinfo{title}{SPIN-ORBIT-COUPLED BOSE-EINSTEIN CONDENSATES}},
  chap. \bibinfo{chapter}{CHAPTER 5}, \bibinfo{pages}{201--250}.
\newblock
  \eprint{https://www.worldscientific.com/doi/pdf/10.1142/9789814667746_0005}.

\bibitem{LinSOC2011}
\bibinfo{author}{Lin, Y.~J.}, \bibinfo{author}{Jim{\'e}nez-Garc{\'\i}a, K.} \&
  \bibinfo{author}{Spielman, I.~B.}
\newblock \bibinfo{journal}{\bibinfo{title}{Spin--orbit-coupled bose--einstein
  condensates}}.
\newblock {\emph{\JournalTitle{Nature}}} \textbf{\bibinfo{volume}{471}},
  \bibinfo{pages}{83--86}, \doiprefix\url{10.1038/nature09887}
  (\bibinfo{year}{2011}).

\bibitem{yunspectrum2013}
\bibinfo{author}{Li, Y.}, \bibinfo{author}{Martone, G.~I.},
  \bibinfo{author}{Pitaevskii, L.~P.} \& \bibinfo{author}{Stringari, S.}
\newblock \bibinfo{journal}{\bibinfo{title}{Superstripes and the excitation
  spectrum of a spin-orbit-coupled bose-einstein condensate}}.
\newblock {\emph{\JournalTitle{Phys. Rev. Lett.}}}
  \textbf{\bibinfo{volume}{110}}, \bibinfo{pages}{235302},
  \doiprefix\url{10.1103/PhysRevLett.110.235302} (\bibinfo{year}{2013}).

\bibitem{Yunbo2017}
\bibinfo{author}{Chen, L.}, \bibinfo{author}{Pu, H.}, \bibinfo{author}{Yu,
  Z.-Q.} \& \bibinfo{author}{Zhang, Y.}
\newblock \bibinfo{journal}{\bibinfo{title}{Collective excitation of a trapped
  bose-einstein condensate with spin-orbit coupling}}.
\newblock {\emph{\JournalTitle{Phys. Rev. A}}} \textbf{\bibinfo{volume}{95}},
  \bibinfo{pages}{033616}, \doiprefix\url{10.1103/PhysRevA.95.033616}
  (\bibinfo{year}{2017}).

\bibitem{Geier2021}
\bibinfo{author}{Geier, K.~T.}, \bibinfo{author}{Martone, G.~I.},
  \bibinfo{author}{Hauke, P.} \& \bibinfo{author}{Stringari, S.}
\newblock \bibinfo{journal}{\bibinfo{title}{Exciting the goldstone modes of a
  supersolid spin-orbit-coupled bose gas}}.
\newblock {\emph{\JournalTitle{Phys. Rev. Lett.}}}
  \textbf{\bibinfo{volume}{127}}, \bibinfo{pages}{115301},
  \doiprefix\url{10.1103/PhysRevLett.127.115301} (\bibinfo{year}{2021}).

\bibitem{Geier2023}
\bibinfo{author}{Geier, K.~T.}, \bibinfo{author}{Martone, G.~I.},
  \bibinfo{author}{Hauke, P.}, \bibinfo{author}{Ketterle, W.} \&
  \bibinfo{author}{Stringari, S.}
\newblock \bibinfo{title}{Dynamics of stripe patterns in supersolid
  spin-orbit-coupled bose gases}, \doiprefix\url{10.48550/ARXIV.2210.10064}
  (\bibinfo{year}{2022}).

\bibitem{mivehvar2021}
\bibinfo{author}{Mivehvar, F.}, \bibinfo{author}{Piazza, F.},
  \bibinfo{author}{Donner, T.} \& \bibinfo{author}{Ritsch, H.}
\newblock \bibinfo{journal}{\bibinfo{title}{Cavity qed with quantum gases: new
  paradigms in many-body physics}}.
\newblock {\emph{\JournalTitle{Advances in Physics}}}
  \textbf{\bibinfo{volume}{70}}, \bibinfo{pages}{1--153},
  \doiprefix\url{10.1080/00018732.2021.1969727} (\bibinfo{year}{2021}).
\newblock \eprint{https://doi.org/10.1080/00018732.2021.1969727}.

\bibitem{ETH_SS1}
\bibinfo{author}{L{\'{e}}onard, J.}, \bibinfo{author}{Morales, A.},
  \bibinfo{author}{Zupancic, P.}, \bibinfo{author}{Esslinger, T.} \&
  \bibinfo{author}{Donner, T.}
\newblock \bibinfo{journal}{\bibinfo{title}{Supersolid formation in a quantum
  gas breaking a continuous translational symmetry}}.
\newblock {\emph{\JournalTitle{Nature}}} \textbf{\bibinfo{volume}{543}},
  \bibinfo{pages}{87} (\bibinfo{year}{2017}).

\bibitem{ETH_SS2}
\bibinfo{author}{L{\'e}onard, J.}, \bibinfo{author}{Morales, A.},
  \bibinfo{author}{Zupancic, P.}, \bibinfo{author}{Donner, T.} \&
  \bibinfo{author}{Esslinger, T.}
\newblock \bibinfo{journal}{\bibinfo{title}{Monitoring and manipulating higgs
  and goldstone modes in a supersolid quantum gas}}.
\newblock {\emph{\JournalTitle{Science}}} \textbf{\bibinfo{volume}{358}},
  \bibinfo{pages}{1415--1418}, \doiprefix\url{10.1126/science.aan2608}
  (\bibinfo{year}{2017}).
\newblock
  \eprint{https://science.sciencemag.org/content/358/6369/1415.full.pdf}.

\bibitem{Zimmerman2020}
\bibinfo{author}{Schuster, S.~C.}, \bibinfo{author}{Wolf, P.},
  \bibinfo{author}{Ostermann, S.}, \bibinfo{author}{Slama, S.} \&
  \bibinfo{author}{Zimmermann, C.}
\newblock \bibinfo{journal}{\bibinfo{title}{Supersolid properties of a
  bose-einstein condensate in a ring resonator}}.
\newblock {\emph{\JournalTitle{Phys. Rev. Lett.}}}
  \textbf{\bibinfo{volume}{124}}, \bibinfo{pages}{143602},
  \doiprefix\url{10.1103/PhysRevLett.124.143602} (\bibinfo{year}{2020}).

\bibitem{Guo2021}
\bibinfo{author}{Guo, Y.} \emph{et~al.}
\newblock \bibinfo{journal}{\bibinfo{title}{An optical lattice with sound}}.
\newblock {\emph{\JournalTitle{Nature}}} \textbf{\bibinfo{volume}{599}},
  \bibinfo{pages}{211--215}, \doiprefix\url{10.1038/s41586-021-03945-x}
  (\bibinfo{year}{2021}).

\bibitem{Hemmerich2015}
\bibinfo{author}{Klinder, J.}, \bibinfo{author}{Ke\ss{}ler, H.},
  \bibinfo{author}{Bakhtiari, M.~R.}, \bibinfo{author}{Thorwart, M.} \&
  \bibinfo{author}{Hemmerich, A.}
\newblock \bibinfo{journal}{\bibinfo{title}{Observation of a superradiant mott
  insulator in the dicke-hubbard model}}.
\newblock {\emph{\JournalTitle{Phys. Rev. Lett.}}}
  \textbf{\bibinfo{volume}{115}}, \bibinfo{pages}{230403},
  \doiprefix\url{10.1103/PhysRevLett.115.230403} (\bibinfo{year}{2015}).

\bibitem{Pitaevskii1984}
\bibinfo{author}{Pitaevskii, L.~P.}
\newblock \bibinfo{journal}{\bibinfo{title}{Layered structure of superfluid
  helium-4 with supercritical motion}}.
\newblock {\emph{\JournalTitle{JETP Letters}}} \textbf{\bibinfo{volume}{39}},
  \bibinfo{pages}{511--514} (\bibinfo{year}{1984}).

\bibitem{AncilottoLev2005}
\bibinfo{author}{Ancilotto, F.}, \bibinfo{author}{Dalfovo, F.},
  \bibinfo{author}{Pitaevskii, L.~P.} \& \bibinfo{author}{Toigo, F.}
\newblock \bibinfo{journal}{\bibinfo{title}{Density pattern in supercritical
  flow of liquid $^{4}\mathrm{He}$}}.
\newblock {\emph{\JournalTitle{Phys. Rev. B}}} \textbf{\bibinfo{volume}{71}},
  \bibinfo{pages}{104530}, \doiprefix\url{10.1103/PhysRevB.71.104530}
  (\bibinfo{year}{2005}).

\bibitem{levons2012}
\bibinfo{author}{Baym, G.} \& \bibinfo{author}{Pethick, C.~J.}
\newblock \bibinfo{journal}{\bibinfo{title}{Landau critical velocity in weakly
  interacting bose gases}}.
\newblock {\emph{\JournalTitle{Phys. Rev. A}}} \textbf{\bibinfo{volume}{86}},
  \bibinfo{pages}{023602}, \doiprefix\url{10.1103/PhysRevA.86.023602}
  (\bibinfo{year}{2012}).

\bibitem{KdV1895}
\bibinfo{author}{Korteweg, D. D.~J.} \& \bibinfo{author}{de~Vries, D.~G.}
\newblock \bibinfo{journal}{\bibinfo{title}{Xli. on the change of form of long
  waves advancing in a rectangular canal, and on a new type of long stationary
  waves}}.
\newblock {\emph{\JournalTitle{The London, Edinburgh, and Dublin Philosophical
  Magazine and Journal of Science}}} \textbf{\bibinfo{volume}{39}},
  \bibinfo{pages}{422--443}, \doiprefix\url{10.1080/14786449508620739}
  (\bibinfo{year}{1895}).
\newblock \eprint{https://doi.org/10.1080/14786449508620739}.

\bibitem{martone2021}
\bibinfo{author}{Martone, G.~I.}, \bibinfo{author}{Recati, A.} \&
  \bibinfo{author}{Pavloff, N.}
\newblock \bibinfo{journal}{\bibinfo{title}{Supersolidity of cnoidal waves in
  an ultracold bose gas}}.
\newblock {\emph{\JournalTitle{Phys. Rev. Res.}}} \textbf{\bibinfo{volume}{3}},
  \bibinfo{pages}{013143}, \doiprefix\url{10.1103/PhysRevResearch.3.013143}
  (\bibinfo{year}{2021}).

\bibitem{Matsuda1970}
\bibinfo{author}{Matsuda, H.} \& \bibinfo{author}{Tsuneto, T.}
\newblock \bibinfo{journal}{\bibinfo{title}{{Off-Diagonal Long-Range Order in
  Solids*)}}}.
\newblock {\emph{\JournalTitle{Progress of Theoretical Physics Supplement}}}
  \textbf{\bibinfo{volume}{46}}, \bibinfo{pages}{411--436},
  \doiprefix\url{10.1143/PTPS.46.411} (\bibinfo{year}{1970}).
\newblock
  \eprint{https://academic.oup.com/ptps/article-pdf/doi/10.1143/PTPS.46.411/5303982/46-411.pdf}.

\bibitem{kottmann2021}
\bibinfo{author}{Kottmann, K.}, \bibinfo{author}{Haller, A.},
  \bibinfo{author}{Ac\'{\i}n, A.}, \bibinfo{author}{Astrakharchik, G.~E.} \&
  \bibinfo{author}{Lewenstein, M.}
\newblock \bibinfo{journal}{\bibinfo{title}{Supersolid-superfluid phase
  separation in the extended bose-hubbard model}}.
\newblock {\emph{\JournalTitle{Phys. Rev. B}}} \textbf{\bibinfo{volume}{104}},
  \bibinfo{pages}{174514}, \doiprefix\url{10.1103/PhysRevB.104.174514}
  (\bibinfo{year}{2021}).

\bibitem{dickeSS2010-esslinger}
\bibinfo{author}{Baumann, K.}, \bibinfo{author}{Guerlin, C.},
  \bibinfo{author}{Brennecke, F.} \& \bibinfo{author}{Esslinger, T.}
\newblock \bibinfo{journal}{\bibinfo{title}{Dicke quantum phase transition with
  a superfluid gas in an optical cavity}}.
\newblock {\emph{\JournalTitle{Nature}}} \textbf{\bibinfo{volume}{464}},
  \bibinfo{pages}{1301--1306}, \doiprefix\url{10.1038/nature09009}
  (\bibinfo{year}{2010}).

\bibitem{dicke-roton2012}
\bibinfo{author}{Mottl, R.} \emph{et~al.}
\newblock \bibinfo{journal}{\bibinfo{title}{Roton-type mode softening in a
  quantum gas with cavity-mediated long-range interactions}}.
\newblock {\emph{\JournalTitle{Science}}} \textbf{\bibinfo{volume}{336}},
  \bibinfo{pages}{1570--1573}, \doiprefix\url{10.1126/science.1220314}
  (\bibinfo{year}{2012}).
\newblock \eprint{https://www.science.org/doi/pdf/10.1126/science.1220314}.

\bibitem{balibarLayer}
\bibinfo{author}{Noury, A.} \emph{et~al.}
\newblock \bibinfo{journal}{\bibinfo{title}{Layering transition in superfluid
  helium adsorbed on a carbon nanotube mechanical resonator}}.
\newblock {\emph{\JournalTitle{Phys. Rev. Lett.}}}
  \textbf{\bibinfo{volume}{122}}, \bibinfo{pages}{165301},
  \doiprefix\url{10.1103/PhysRevLett.122.165301} (\bibinfo{year}{2019}).

\bibitem{Choi2021}
\bibinfo{author}{Choi, J.}, \bibinfo{author}{Zadorozhko, A.~A.},
  \bibinfo{author}{Choi, J.} \& \bibinfo{author}{Kim, E.}
\newblock \bibinfo{journal}{\bibinfo{title}{Spatially modulated superfluid
  state in two-dimensional $^{4}\mathrm{He}$ films}}.
\newblock {\emph{\JournalTitle{Phys. Rev. Lett.}}}
  \textbf{\bibinfo{volume}{127}}, \bibinfo{pages}{135301},
  \doiprefix\url{10.1103/PhysRevLett.127.135301} (\bibinfo{year}{2021}).

\bibitem{physLayerSS}
\bibinfo{author}{Rao, R.}
\newblock \bibinfo{journal}{\bibinfo{title}{Seeking supersolidity in helium
  layers}}.
\newblock {\emph{\JournalTitle{Physics}}} \textbf{\bibinfo{volume}{14}},
  \bibinfo{pages}{s114} (\bibinfo{year}{2021}).

\bibitem{Radzi2009}
\bibinfo{author}{Radzihovsky, L.} \& \bibinfo{author}{Vishwanath, A.}
\newblock \bibinfo{journal}{\bibinfo{title}{Quantum liquid crystals in an
  imbalanced fermi gas: Fluctuations and fractional vortices in
  larkin-ovchinnikov states}}.
\newblock {\emph{\JournalTitle{Phys. Rev. Lett.}}}
  \textbf{\bibinfo{volume}{103}}, \bibinfo{pages}{010404},
  \doiprefix\url{10.1103/PhysRevLett.103.010404} (\bibinfo{year}{2009}).

\bibitem{casalbuoni2004}
\bibinfo{author}{Casalbuoni, R.} \& \bibinfo{author}{Nardulli, G.}
\newblock \bibinfo{journal}{\bibinfo{title}{Inhomogeneous superconductivity in
  condensed matter and qcd}}.
\newblock {\emph{\JournalTitle{Rev. Mod. Phys.}}}
  \textbf{\bibinfo{volume}{76}}, \bibinfo{pages}{263--320},
  \doiprefix\url{10.1103/RevModPhys.76.263} (\bibinfo{year}{2004}).

\bibitem{yang2001}
\bibinfo{author}{Yang, K.}
\newblock \bibinfo{journal}{\bibinfo{title}{Inhomogeneous superconducting state
  in quasi-one-dimensional systems}}.
\newblock {\emph{\JournalTitle{Phys. Rev. B}}} \textbf{\bibinfo{volume}{63}},
  \bibinfo{pages}{140511}, \doiprefix\url{10.1103/PhysRevB.63.140511}
  (\bibinfo{year}{2001}).

\bibitem{orso2006}
\bibinfo{author}{Orso, G.}
\newblock \bibinfo{journal}{\bibinfo{title}{Attractive fermi gases with unequal
  spin populations in highly elongated traps}}.
\newblock {\emph{\JournalTitle{Phys. Rev. Lett.}}}
  \textbf{\bibinfo{volume}{98}}, \bibinfo{pages}{070402},
  \doiprefix\url{10.1103/PhysRevLett.98.070402} (\bibinfo{year}{2007}).

\bibitem{Hulet2010}
\bibinfo{author}{Liao, Y.-a.} \emph{et~al.}
\newblock \bibinfo{journal}{\bibinfo{title}{Spin-imbalance in a one-dimensional
  fermi gas}}.
\newblock {\emph{\JournalTitle{Nature}}} \textbf{\bibinfo{volume}{467}},
  \bibinfo{pages}{567--569}, \doiprefix\url{10.1038/nature09393}
  (\bibinfo{year}{2010}).

\bibitem{pavani2022}
\bibinfo{author}{Pavarini, E.}
\newblock \bibinfo{journal}{\bibinfo{title}{Superconductors gain momentum}}.
\newblock {\emph{\JournalTitle{Science}}} \textbf{\bibinfo{volume}{376}},
  \bibinfo{pages}{350--351}, \doiprefix\url{10.1126/science.abn3794}
  (\bibinfo{year}{2022}).
\newblock \eprint{https://www.science.org/doi/pdf/10.1126/science.abn3794}.

\bibitem{FFLO-organic}
\bibinfo{author}{Wosnitza, J.}
\newblock \bibinfo{journal}{\bibinfo{title}{Spatially nonuniform
  superconductivity in quasi-two-dimensional organic charge-transfer salts}}.
\newblock {\emph{\JournalTitle{Crystals}}} \textbf{\bibinfo{volume}{8}},
  \doiprefix\url{10.3390/cryst8050183} (\bibinfo{year}{2018}).

\bibitem{chamel2008}
\bibinfo{author}{Chamel, N.} \& \bibinfo{author}{Haensel, P.}
\newblock \bibinfo{journal}{\bibinfo{title}{Physics of neutron star crusts}}.
\newblock {\emph{\JournalTitle{Living Reviews in Relativity}}}
  \textbf{\bibinfo{volume}{11}}, \bibinfo{pages}{10},
  \doiprefix\url{10.12942/lrr-2008-10} (\bibinfo{year}{2008}).

\bibitem{WatPethick2017}
\bibinfo{author}{Watanabe, G.} \& \bibinfo{author}{Pethick, C.~J.}
\newblock \bibinfo{journal}{\bibinfo{title}{Superfluid density of neutrons in
  the inner crust of neutron stars: New life for pulsar glitch models}}.
\newblock {\emph{\JournalTitle{Phys. Rev. Lett.}}}
  \textbf{\bibinfo{volume}{119}}, \bibinfo{pages}{062701},
  \doiprefix\url{10.1103/PhysRevLett.119.062701} (\bibinfo{year}{2017}).

\end{thebibliography}

\section*{Acknowledgements}

We acknowledge fruitful and stimulating discussions and collaborations  during the last few years with Francesca Ferlaino, Albert Gallem\'i, Kevin Geier, Philipp Hauke, Wolfgang Ketterle, Giovanni Martone, Giovanni Modugno, Tilman Pfau, Francesco Piazza, Santo Roccuzzo, Luis Santos and Leticia Tarruell.

We acknowledge funding from Provincia Autonoma di Trento, from the Italian MIUR under the PRIN2017 project CEnTraL (Protocol Number 20172H2SC4), and from PNRR MUR project PE0000023-NQSTI.




\end{document}